\documentclass[11pt]{article}

\usepackage[margin=1in]{geometry}
\usepackage{amsmath,amssymb,amsthm,mathtools}
\usepackage{enumitem}
\usepackage{aliascnt}
\usepackage{xcolor}
\usepackage{hyperref}
\usepackage{cleveref}
\usepackage{microtype}

\newtheorem{Thm}{Theorem}[section]
\newtheorem*{Thm*}{Theorem}

\newaliascnt{Prop}{Thm}
\newtheorem{Prop}[Prop]{Proposition}
\aliascntresetthe{Prop}

\newaliascnt{Lem}{Thm}
\newtheorem{Lem}[Lem]{Lemma}
\aliascntresetthe{Lem}

\newaliascnt{Cor}{Thm}
\newtheorem{Cor}[Cor]{Corollary}
\aliascntresetthe{Cor}

\newaliascnt{Fact}{Thm}
\newtheorem{Fact}[Fact]{Fact}
\aliascntresetthe{Fact}

\newtheorem*{Con*}{Condition}

\theoremstyle{definition}
\newtheorem{construction}{Construction}[section]
\crefname{construction}{Construction}{Constructions}

\crefname{Thm}{Theorem}{Theorems}
\crefname{Prop}{Proposition}{Propositions}
\crefname{Lem}{Lemma}{Lemmas}
\crefname{Fact}{Fact}{Facts}
\crefname{Def}{Definition}{Definitions}
\crefname{Cor}{Corollary}{Corollaries}
\crefname{Con}{Condition}{Conditions}
\crefname{equation}{Equation}{Equations}
\crefname{section}{Section}{Sections}
\crefname{appendix}{Appendix}{Appendices}

\newcommand{\IC}{\operatorname{IC}}
\newcommand{\Cost}{\operatorname{Cost}}
\newcommand{\err}{\operatorname{err}}
\newcommand{\TV}{\operatorname{TV}}
\newcommand{\E}{\mathbb E}
\newcommand{\Mand}{\text{~and~}}
\newcommand{\supp}{\mathrm{supp}}
\newcommand{\muF}{{\mu_\mathrm{full}}}
\renewcommand{\Pr}{\mathop{\mathrm{Pr}}\nolimits}
\newcommand{\proofstep}[1]{\par\medskip\noindent\textit{#1}\enspace}
\title{The Exact Communication Threshold for Parallel Repetition}
\author{Daiki Suruga\\IQC, University of Waterloo}
\date{}
\begin{document}
\maketitle
\begin{abstract}
The parallel repetition theorem is a fundamental tool in theoretical computer science, showing that the probability of winning all $n$ independent instances of a finite two-player game decays exponentially in $n$, provided that the game has no perfect strategy.
The classical theorem assumes that the players cannot communicate, raising the question of how much communication can be allowed while preserving this exponential decay.

For every finite total relation $g$ and arbitrary input distribution $\mu$, we prove exponential decay at every fixed communication rate below a quantity called zero-error internal information complexity, denoted by $\mathrm{IC}_{\mu}(g,0)$.
Together with known results showing that $n\,\mathrm{IC}_{\mu}(g,0)+o(n)$ bits of communication suffice to solve all $n$ instances with success probability tending to one, our bound determines the exact communication threshold.
As a by-product, we obtain exact characterizations of amortized distributional and randomized communication complexity under the global-error criterion.
\end{abstract}

\begingroup
\renewcommand{\abstractname}{AI disclosure}
\begin{abstract}
The author used ChatGPT (OpenAI) to discover the main result and assist with literature searches and the preparation and revision of this manuscript. Access was provided through OpenAI's ChatGPT for Academic Researchers program. 
For transparency, an earlier draft of this manuscript, written by ChatGPT, is available on the \href{https://osf.io/jpyhn/overview?view_only=8458b1310df3481f9f0fed7476597443}{Open Science Framework (OSF)}.
The author reviewed and improved the material incorporated into the manuscript and takes full responsibility for its content.
\end{abstract}
\endgroup

\section{Introduction}
\label{sec:introduction}
For a finite two-player game with a total relation $g$
and question distribution $\mu$, let $\omega(g,\mu)$ denote its
largest winning probability without communication.
The classical parallel repetition theorem of
Refs.~\cite{Raz98,Hol07} states that, if $\omega(g,\mu)<1$, the
largest probability of winning all $n$ independent instances $(g^n, \mu^{\otimes n})$
satisfies
\[
  \omega(g^n,\mu^{\otimes n})\le 2^{-\alpha n}
\]
for some constant $\alpha=\alpha(g,\mu)>0$.
Parallel repetition theorems have diverse applications: they are used to reduce soundness error in two-prover interactive proofs~\cite{Raz98} and play a central role in hardness-of-approximation results for constraint satisfaction, Clique, and Set Cover~\cite{Has01,Fei98,DS14} for example.

This paper studies how this exponential decay changes when Alice
and Bob are allowed to communicate after receiving their question
strings.
Allowing enough communication eliminates the decay altogether:
exchanging the full inputs guarantees success because $g$ is total.

The classical theorems already give a partial answer: exponential
decay persists when the communication is sufficiently
small.
Indeed, for a protocol $\Pi_n$ for $g^n$ communicating at most
$c_n$ bits, guessing its transcript using shared randomness gives
a zero-communication strategy whose success probability is at least
$2^{-c_n} \cdot \Pr[\Pi_n\text{ succeeds}]$; see, e.g., the transcript guessing argument~\cite[Section 4.1]{JK22}.
The repeated-game value and the preceding classical bound therefore
give
\[
  \Pr_{\mu^{\otimes n}}[\Pi_n\text{ succeeds}]
  \le
  2^{c_n}\omega(g^n,\mu^{\otimes n})
  \le
  2^{-(n\alpha-c_n)}.
\]
Hence, the success probability of the protocol $\Pi_n$ decays
exponentially whenever $c_n\ll\alpha n$.
However, $\alpha$ can be very small; $\alpha<1$ whenever
$\omega(g,\mu)>1/2$,\footnote{Observe that
$\omega(g,\mu)^n\le\omega(g^n,\mu^{\otimes n})\le2^{-\alpha n}$.}
and even the exponent supplied by the current best bound~Ref.~\cite{Hol07} is also less than one.
When $\alpha<1$, the condition $c_n\ll\alpha n$ requires
$c_n/n\ll1$, and therefore, the communication allowance obtained by this
argument is below one bit per instance.
This restriction to less than one bit can be far
from optimal: for some functions $g$ on $m$-bit inputs and
suitable input distributions, the success probability is known
to decay exponentially in $n$ even with $\Omega(nm)$ bits of
communication; see Ref.~\cite[Corollary 23]{LSS08}.

To fill this gap, this paper investigates the exact communication
threshold:
\begin{quote}
\emph{
how much communication can be allowed while every
protocol still has exponentially small success probability,
and how much suffices for success tending to one?
}
\end{quote}
This extends the question addressed by classical parallel repetition,
which is recovered when $c_n=0$.

This paper determines this exact threshold on the asymptotic rate $\lim_n c_n/n$ for every finite total relation $g$
and arbitrary input distribution $\mu$: it equals
$\mathrm{IC}_\mu(g,0)$, the zero-error value of a quantity called
information complexity (see Ref.~\cite{BR11}).
Formally, for every $\delta > 0$, we prove that every two-party protocol $\Pi_n$ having the worst-case cost $|\Pi_n|$ at most $n(\IC_\mu(g, 0) - \delta)$
succeeds on $(g^n,\mu^{\otimes n})$ with probability at most $2^{-\Omega( n)}$.
Conversely, the compression result of Ref.~\cite{BR11} gives protocols $\Pi_n$ with communication cost $n(\IC_\mu(g, 0) + o(1))$ that succeed with probability tending to one.
Together, these bounds also determine the exact asymptotic distributional
communication complexity of $g^n$ for every global error
$\varepsilon \in (0,1)$; we additionally obtain the corresponding characterization for randomized communication complexity.
Our result therefore refines classical parallel repetition by
identifying the exact communication threshold below which exponential
decay persists.
\cref{section:our-results:intro} gives the precise statements and consequences.
\subsection{Our results and consequences}\label{section:our-results:intro}
To state our main result precisely, we first introduce the setting
and notation.
Let $g\subseteq\mathcal X\times\mathcal Y\times\mathcal A\times\mathcal B$
be a relation on finite sets, where $\mathcal X \times \mathcal Y$ is the set of inputs, and $\mathcal A\times\mathcal B$ is the set of outputs.
(We assume that $g$ is total, meaning that a valid output pair exists
for every input pair.)
Let $\mu$ be an arbitrary distribution on
$\mathcal X\times\mathcal Y$, and write $\mu^{\otimes n}$ for its $n$-i.i.d.
distribution.
We consider general two-party protocols with an arbitrary number of rounds and measure their
worst-case communication cost, formally given in~\cref{sec:setting}.

Let us also briefly introduce a quantity that plays a crucial
role in our result.
This quantity, called internal information complexity, is an
information-theoretic measure used in communication complexity;
see Refs.~\cite{BR11,Bra17,RY20}.
Specifically, our main result uses the zero-error internal
information complexity of $g$ under the input distribution $\mu$,
denoted by $\mathrm{IC}_\mu(g,0)$.
The formal definitions are given in~\cref{sec:setting}.

Our main result is the following, which is a simplified version of \cref{thm:converse};
the full statement also gives explicit bounds on the exponent $\kappa$.
\begin{Thm}[Main result]\label{main_theorem:intro}
For every $0\le R<\mathrm{IC}_\mu(g,0)$, there exists
$\kappa=\kappa(g,\mu,R)>0$ such that, for every $n\ge 1$, every
 protocol $\Pi_n$ for $g^n$ communicating at most
$nR$ bits satisfies
\[
  \Pr_{\mu^{\otimes n}}
  \bigl[\Pi_n\text{ succeeds on every coordinate}\bigr]
  \le 2^{-\kappa n}.
\]
\end{Thm}

Thus, every communication rate below
$\mathrm{IC}_\mu(g,0)$ forces exponentially small success probability,
with an exponent independent of the protocol and the number of
instances.
Consequently, even protocols with success probability $2^{-o(n)}$
require at least
$n\bigl(\mathrm{IC}_\mu(g,0)-o(1)\bigr)$ bits of worst-case
communication.
Moreover, at zero communication, our explicit exponent for $\kappa$ recovers the current best bound of Ref.~\cite{Hol07}.

\paragraph{Consequences.} 
\cref{main_theorem:intro} has several consequences.
The first consequence determines the exact asymptotic distributional communication
complexity of a finite total relation $g$.
For a fixed $0<\varepsilon<1$, let
$D_{\mu^{\otimes n},\varepsilon}(g^n)$ denote the distributional
communication complexity of $g^n$ under $\mu^{\otimes n}$ with error
at most $\varepsilon$; see \cref{sec:setting} for the formal definition.
\cref{main_theorem:intro} rules out every fixed rate
below $\mathrm{IC}_\mu(g,0)$, since the success probability
eventually falls below $1-\varepsilon$.
Conversely, applying the compression result of Ref.~\cite{BR11}
to independent executions of a zero-error protocol achieves
every fixed rate above $\mathrm{IC}_\mu(g,0)$ with vanishing
overall error.
Therefore,
\begin{equation}\label{intro:eq:distri:am}
  \lim_{n\to\infty}
  \frac{D_{\mu^{\otimes n},\varepsilon}(g^n)}{n}
  =\mathrm{IC}_\mu(g,0).
\end{equation}
Thus, the asymptotic communication rate is independent of the fixed
error allowance $\varepsilon\in(0,1)$ and is determined by the
zero-error information complexity.
\cref{thm:distributional-amortization} gives the full statement and proof.

Our result also has consequences for randomized communication complexity.
Let $\mathrm{IC}(g,0)$ denote the maximum of $\IC_\mu(g,0)$ over all input distributions $\mu$.
By definition, for every $0\le R<\mathrm{IC}(g,0)$, there is an input
distribution $\mu$ such that $R<\mathrm{IC}_\mu(g,0)$.
Thus, applying the main theorem then gives exponential decay of the success probability when inputs are averaged over $\mu^{\otimes n}$, which gives the desired exponential decay of success probability on at least one input.
See \cref{cor:prior-free-strong-converse} for a formal statement.

Additionally, combining this exponential decay with the upper bound of Ref.~\cite{Sur26} determines the asymptotic randomized communication rate exactly.
For a fixed $0<\varepsilon<1$, let $R_\varepsilon(g^n)$ denote
the randomized communication complexity of $g^n$ with error
at most $\varepsilon$, as defined in \cref{sec:setting}.
As in the derivation of Equality~\eqref{intro:eq:distri:am}, the implication for the randomized communication rules out every fixed rate below $\mathrm{IC}(g,0)$ for all sufficiently large $n$.
Together with the upper bound~\cite{Sur26}, this establishes
\[
  \lim_{n\to\infty}\frac{R_\varepsilon(g^n)}{n}
  =\mathrm{IC}(g,0).
\]
See \cref{thm:prior-free-amortization} for the full statement and proof.

\subsection{Prior works}\label{section:intro:Prior}
We compare our results with earlier communication direct-product theorems, strong converses, and characterizations of amortized communication complexity, explaining which communication thresholds were previously known and which are determined by our theorem.

\paragraph{Direct product theorems.}
We first compare our result with earlier strong\footnote{A direct-product theorem is called strong if, writing $C$ for the communication complexity of one instance at constant error, it asserts that every protocol using at most $cnC$ bits has probability at most $2^{-\Omega(n)}$ of solving all $n$ instances correctly, for some constant $c>0$. A weak direct-product theorem allows only communication comparable to $C$.} direct-product theorems~\cite{BRWY13,JPY12,BW15,Yu22,IR24} in communication complexity.
For example, for a Boolean function $g$ under an arbitrary input
distribution $\mu$, Ref.~\cite{BRWY13} proves exponentially small
success whenever $|\Pi_n|\ge 2$ and
\[
  |\Pi_n|(\log|\Pi_n|)^{3/2}
  < c\bigl(D_\mu(g,1/3)-2\bigr)\sqrt n,
\]
 where $|\Pi_n|$ is the worst-case communication cost of $\Pi_n$
and $c>0$ is a sufficiently small universal constant.
More recently, Ref.~\cite{IR24} proves stronger results, known as XOR
lemmas, and derives corresponding direct-product bounds, with a similar $\sqrt n$ dependence on the
number of instances up to logarithmic factors.

As a prior work using internal information complexity $\IC_\mu(g,1/3)$ with error $\le 1/3$, Ref.~\cite{BW15} proves
exponentially small success probability whenever $|\Pi_n|\ge 2$ and
\[
  |\Pi_n|\log\!\bigl(|\Pi_n|\log|\Pi_n|\bigr)
  \le c n\,\mathrm{IC}_\mu(g,1/3),
\]
again for a sufficiently small universal constant $c>0$. Ref.~\cite{BK18_repetition} also discusses the relation between parallel repetition and information complexity but without communication.

These results do not determine the optimal communication threshold
for exponential decay for general tasks.
In contrast, \cref{main_theorem:intro} establishes exponential decay
at every fixed rate below $\mathrm{IC}_\mu(g,0)$ for every finite total
relation $g$ and arbitrary input distribution $\mu$, giving the exact threshold for the first time together with the matching upper bound discussed above.

Additionally, combining our theorem with the lower bounds on $\IC_\mu(g,0)$ of Refs.~\cite{BJKS04,JKS03,BW16} recovers several strong direct-product conclusions.
Examples include those for Set Disjointness, Tribes, and Inner Product previously obtained in Refs.~\cite{Kla10,HJ13,LSS08}, respectively.

\paragraph{Works from information theory.}
Our result is also closely related to work in information theory,
where such exponential decay of success is studied under the name
of \emph{exponential strong converses}.
For a function $g$, Ref.~\cite{TVVW17} proves, among other results,
that for every fixed $\delta>0$, the success probability of $\Pi_n$
is exponentially small whenever
\[
  |\Pi_n|
  \le n\bigl[H(g(X,Y)\mid X)+H(g(X,Y)\mid Y)-\delta\bigr],
\]
where $(X,Y)\sim\mu$ and $H(\cdot\mid\cdot)$ denotes conditional
entropy.
Notably, Ref.~\cite{TVVW17} explicitly identifies replacing this
entropy threshold by zero-error information complexity
$\mathrm{IC}_\mu(g,0)$ as the desired strengthening.
The function case of \cref{main_theorem:intro} achieves this
strengthening.

Refs.~\cite{TW20,HSW24} provide further close comparisons:
they establish exponential strong converses \emph{with a fixed number} of communication rounds, at the corresponding
round-restricted rates.
Our result has no restriction on the number of rounds.

\paragraph{Amortized communication complexity.}
Information complexity also characterizes amortized communication
under \emph{a per-coordinate error requirement}.
Ref.~\cite{BR11} shows that $\mathrm{IC}_\mu(g,\varepsilon)$
equals the amortized distributional communication complexity
when the error probability on each coordinate is at most
$\varepsilon$.
Ref.~\cite{Bra17} establishes the corresponding 
characterization of randomized communication in terms of $\mathrm{IC}(g,\varepsilon)$.
Under a strict requirement that all coordinates be correct
simultaneously, Ref.~\cite{Sur26} proves that,
for every fixed $0<\varepsilon<1$, the amortized rate is exactly
$(1-\varepsilon)\mathrm{IC}(g,0)$ when communication cost is measured
in expectation.
For worst-case randomized communication complexity, Ref.~\cite{Sur26} establishes
the bounds
\[
  (1-\varepsilon)\mathrm{IC}(g,0)
  \le \liminf_{n\to\infty}\frac{R_\varepsilon(g^n)}n
  \le \limsup_{n\to\infty}\frac{R_\varepsilon(g^n)}n
  \le \mathrm{IC}(g,0).
\]
Our result closes this interval at its upper bound 
and, more strongly, proves exponential decay of success probability
at every rate below that value.

\subsection{Proof techniques and comparison}\label{section:intro:Proof}
We briefly outline our proof strategy to prepare for the comparison
with prior techniques below; \cref{sec:main-result} gives a detailed proof
overview using the terminology and notation introduced in
\cref{sec:setting}.

\paragraph{Proof technique.} 
The proof proceeds by contradiction.
Fix $0\le R<\mathrm{IC}_\mu(g,0)$, and suppose that an
$n$-instance protocol $\Pi_n$ communicates at most $nR$ bits
and succeeds too often.
We will use $\Pi_n$ to construct, for a suitable $\varepsilon>0$,
a protocol for $g$ with error at most $\varepsilon$ and
information cost strictly below $\mathrm{IC}_\mu(g,\varepsilon)$,
contradicting its definition.
For this construction, we choose $\varepsilon>0$ such that
$R<\mathrm{IC}_\mu(g,\varepsilon)$, which is possible by
continuity of $\IC_\mu(g, \varepsilon)$ at zero error.

We first construct a single-instance distribution
$q_{\mathrm{single}}$ from $\Pi_n$ as follows.
Let $P_{\Pi_n}$ denote the joint distribution of
\[
  \bigl(
    \underbrace{(X^n,Y^n)}_{\text{inputs}},
    \underbrace{(A^n,B^n)}_{\text{outputs}},
    \underbrace{T_n}_{\text{transcript}}
  \bigr)
\]
induced by $\Pi_n$ under $\mu^{\otimes n}$.
For a suitably chosen set $S\subseteq[n]$, define the event
\[
  W_S=\bigcap_{i\in S}\{(A_i,B_i)\in g(X_i,Y_i)\}.
\]
Let us sample the inputs, outputs, and transcript according to
the distribution $P_{\Pi_n}(\cdot\mid W_S)$ after conditioning on $W_S$, and independently choose
$J\sim\mathrm{Unif}([n])$.
Let $q_{\mathrm{single}}$ be the resulting joint distribution
of $(X,Y,A,B,V)$, where
\[
  (X,Y,A,B)=(X_J,Y_J,A_J,B_J)
\]
and $V$ is the auxiliary variable defined in
\cref{sec:main-result}.
The set $S$ and the variables retained in $V$ are chosen
to ensure two guarantees:
\textup{(i)}
$q_{\mathrm{single}}[(A,B)\notin g(X,Y)]\le\varepsilon/2$,
and \textup{(ii)} the information value
\[
  I_{q_{\mathrm{single}}}(X;V\mid Y)
  +I_{q_{\mathrm{single}}}(Y;V\mid X)
\]
is at most $R$ plus a small additive loss.

To derive the contradiction from guarantees \textup{(i)} and
\textup{(ii)}, we need corresponding bounds for a protocol
on inputs drawn from $\mu$.
The distribution $q_{\mathrm{single}}$ need not arise from
such a protocol: conditioning may change the input marginal,
and $V$ need not be a valid protocol transcript.
To this end, we construct a desired protocol $\pi_{\mathrm{single}}$
using the conditional distributions of $q_{\mathrm{single}}$,
with each message generated using only the sender's input
and the preceding shared information, as follows.

Write $V=(V_{\mathrm{pub}},V_1,\ldots,V_m)$.
On inputs $(X,Y)\sim\mu$, the players first sample
$V_{\mathrm{pub}}$ publicly according to
$q_{\mathrm{single}}(V_{\mathrm{pub}})$, independently of
their inputs, and then alternately generate the messages
$V_1,\ldots,V_m$.
The resulting joint distribution is
\begin{equation}\label{eq:single-protocol}
\begin{aligned}
  P_{\pi_{\mathrm{single}}}(X,Y,A,B,V)
  &=
  \mu(X,Y)\,q_{\mathrm{single}}(V_{\mathrm{pub}}) \\
  &\quad{}\times
  \underbrace{
    q_{\mathrm{single}}(V_1\mid X,V_{\mathrm{pub}})
  }_{\text{Alice}}
  \underbrace{
    q_{\mathrm{single}}(V_2\mid Y,V_{\mathrm{pub}},V_1)
  }_{\text{Bob}} \\
  &\quad{}\times
  \underbrace{
    q_{\mathrm{single}}(V_3\mid X,V_{\mathrm{pub}},V_1,V_2)
  }_{\text{Alice}}
  \cdots \\
  &\quad{}\times
  q_{\mathrm{single}}(A\mid X,V)\,
  q_{\mathrm{single}}(B\mid Y,V).
\end{aligned}
\end{equation}
Each message and output is sampled using only information
available to the respective player, so this defines a valid
protocol.

We now compare $P_{\pi_{\mathrm{single}}}$ with
$q_{\mathrm{single}}$.
Write $\delta=\operatorname{TV}\bigl(P_{\pi_{\mathrm{single}}},q_{\mathrm{single}}\bigr).$
Guarantee \textup{(i)} then gives
\[
\begin{aligned}
  \operatorname{err}_\mu(\pi_{\mathrm{single}})
  &=
  P_{\pi_{\mathrm{single}}}[(A,B)\notin g(X,Y)] \\
  &\le
  q_{\mathrm{single}}[(A,B)\notin g(X,Y)]+\delta \\
  &\le \varepsilon/2+\delta.
\end{aligned}
\]
Using guarantee \textup{(ii)} and continuity of conditional
mutual information, we also obtain
\[
\begin{aligned}
  \mathrm{IC}_\mu(\pi_{\mathrm{single}})
  &=
  I_{P_{\pi_{\mathrm{single}}}}(X;V\mid Y)
  +I_{P_{\pi_{\mathrm{single}}}}(Y;V\mid X) \\
  &\le R+\eta,
\end{aligned}
\]
where $\eta\ge0$ accounts for both the additive loss in
\textup{(ii)} and the change in the information expression
when replacing $q_{\mathrm{single}}$ by
$P_{\pi_{\mathrm{single}}}$.
We prove that, when $\Pi_n$ succeeds sufficiently often,
these losses are small enough to satisfy
\[
  \delta\le\varepsilon/2,
  \qquad
  \eta<\mathrm{IC}_\mu(g,\varepsilon)-R.
\]
The resulting protocol therefore has error at most
$\varepsilon$ and information cost strictly below
$\mathrm{IC}_\mu(g,\varepsilon)$, giving the desired contradiction.

\paragraph{Comparison to prior work.}
The above strategy of conditioning on events such as success on selected coordinates,
followed by single-coordinate extraction, is an established approach
in parallel repetition and communication direct-product proofs;
see Refs.~\cite{Hol07,BRWY13} for example.
In related work on information and communication complexity,
Refs.~\cite{Yu22,IR24,SY25} obtain intermediate single-instance
distributions playing a role similar to $q_{\mathrm{single}}$
through conditioning or decomposition of the original protocol $\Pi_n$, while controlling their
error together with appropriate information quantities.
For our contradiction, the extracted distribution
$q_{\mathrm{single}}$ must simultaneously have small error and satisfy
\[
  I_{q_{\mathrm{single}}}(X;V\mid Y)
  +I_{q_{\mathrm{single}}}(Y;V\mid X)
  \le R+o(1),
\]
where the additive loss vanishes as $-\log(\Pr[\Pi_n~\text{succeeds}])/n\to0$. 
For comparison, the corresponding lemma of
Ref.~\cite{SY25} gives a single-instance information bound of
$O(R+1)$ when the $n$-instance protocol has information cost
at most $nR$, without requiring the coefficient of $R$ to be one.
Our extraction preserves the coefficient one on $R$ and incurs
only a vanishing additive loss.

Prior works have also developed techniques for obtaining valid
protocols from conditioned distributions; see
Refs.~\cite{BRWY13,BW15,TW20,SY25}.
Our construction of $\pi_{\mathrm{single}}$ from
$q_{\mathrm{single}}$ in~\eqref{eq:single-protocol}
uses local conditional distributions, as in Ref.~\cite{SY25}.

\subsection{Organization of the paper}
The remainder of the paper is organized as follows.
\begin{description}
\item[\Cref{sec:setting}] introduces the communication model and information measures and provides preliminary facts.
\item[\Cref{sec:main-result}] states our main theorem, \cref{thm:converse}, and gives an overview of its proof.
\item[\Cref{sec:reduction,sec:fresh-input,sec:zero-communication,sec:proof_SC}] prove \cref{thm:converse}, following the overview in~\cref{sec:overview}.
\item[\Cref{sec:consequences}] derives the exact asymptotic communication rates and corresponding strong converses in the distributional and randomized settings from~\cref{thm:converse}.
\item[Appendices~\ref{app:facts}, \ref{app:zero-error}, and~\ref{app:conditional-sampling}] provide supplementary results needed for the proof.
\end{description}

\section{Model and information measures}
\label{sec:setting}

\subsection{Relations and protocols}\label{sec:Relations}
Our result compares communication cost for $g^n$ with the
information cost of one copy of $g$. We first specify what it means
to solve the relation and then define these two costs.

Let $\mathcal X,\mathcal Y,\mathcal A,\mathcal B$ be nonempty finite
sets. Alice receives $x\in\mathcal X$ and Bob receives
$y\in\mathcal Y$. They output $a\in\mathcal A$ and
$b\in\mathcal B$ locally. The relation
$g\subseteq\mathcal X\times\mathcal Y\times\mathcal A\times\mathcal B$
is total: for each $(x,y)$, at least one output pair is valid,
\[
 g(x,y):=\{(a,b):(x,y,a,b)\in g\}\ne\varnothing.
\]
On $n$ input pairs, success for $g^n$ means that
$(A_i,B_i)\in g(X_i,Y_i)$ for every $i\in[n]$. We draw
$(X_1,Y_1),\ldots,(X_n,Y_n)$ independently from one fixed distribution
$\mu$ on $\mathcal X\times\mathcal Y$. 

We consider finite classical interactive protocols with binary
communication. The sender of each bit is determined by the public
history, consisting of the transcript and shared randomness. 
The inputs, Public randomness $R_\textrm{pub}$, Alice's private randomness, and
Bob's private randomness are all mutually independent. 
Each output depends only on that player's input,
its private randomness, and the public history. All random
variables have finite alphabets. The worst-case number of communicated bits in a protocol $\pi$ is
$|\pi| = \Cost_{\mathrm{wc}}(\pi)$.

For a function $f:\mathcal X\times\mathcal Y\to\mathcal Z$ considered for communication complexity applications, we use
the relation
\[
 g_f=\{(x,y,a,b):f(x,y)=a=b\}
 \subseteq\mathcal X\times\mathcal Y\times\mathcal Z\times\mathcal Z.
\]
Thus both parties must output the value of $f$.
\subsection{Communication and information measures}

In the standard manner, for $0\le\varepsilon\le1$, define
\begin{align}
	D_{\mu^{\otimes n},\varepsilon}(g^n)
 &:=\inf_{\Pi:\,\Pr_{\mu^{\otimes n}}[\Pi\text{ fails on some coordinate}]
                \le\varepsilon}\Cost_{\mathrm{wc}}(\Pi),
 \label{eq:distributional-communication}\\
 R_\varepsilon(g^n)
 &:=\inf_{\Pi:\,\max_{(x^n,y^n)}
       \Pr[\Pi\text{ fails on some coordinate}\mid x^n,y^n]
                \le\varepsilon}\Cost_{\mathrm{wc}}(\Pi).
 \label{eq:prior-free-communication}
\end{align}
The probability conditional on a fixed input is over the protocol's
randomness. 

For a single-instance protocol $\pi$ on $(X,Y)\sim\mu$, let $T$
denote its communicated transcript and let $P_\pi$ be the joint
distribution of its inputs, public randomness $R_\mathrm{pub}$, transcript, and
outputs $AB$. Define its error and information cost by
\begin{align}
 \err_\mu(\pi)&=\Pr_{P_\pi}[(A,B)\notin g(X,Y)],\notag\\
 \IC_\mu(\pi)&=I_{P_\pi}(X;T\mid Y,R_\mathrm{pub})
                    +I_{P_\pi}(Y;T\mid X,R_\mathrm{pub}),
 \label{eq:information-cost}
\end{align}
where $I_P(X;Y \mid Z)$ is the conditional mutual information under the distribution $P$.
We also define
\begin{equation}
 \IC_\mu(g,\varepsilon)
 =\inf_{\pi:\,\err_\mu(\pi)\le\varepsilon}\IC_\mu(\pi),
\qquad
 \err_\mu(R)
=\inf_{\pi:\,\IC_\mu(\pi)\le R}\err_\mu(\pi)
 \qquad(R\ge0).
 \label{eq:def_IC_error}
\end{equation}

For a non-distributional information measure, require the protocol to meet
the error bound on every input and take its largest information cost
over input distributions:
\begin{equation}
 \IC(g,\varepsilon)
 =\inf_{\pi:\,\max_{(x,y)}\Pr[\pi\text{ errs}\mid x,y]
                         \le\varepsilon}\max_{\nu}\IC_\nu(\pi).
 \label{eq:prior-free-information-complexity}
\end{equation}
Here $\nu$ ranges over distributions on $\mathcal X\times\mathcal Y$.

For the same distribution $\mu$, the best success probability without
communication and its gap from one are
\begin{equation}
 \omega(g,\mu)
 =\max_{a:\mathcal X\to\mathcal A,\ b:\mathcal Y\to\mathcal B}
    \Pr_\mu[(a(X),b(Y))\in g(X,Y)],
 \qquad \eta=1-\omega(g,\mu).
 \label{eq:game-value}
\end{equation}

\subsection{Probability and string notation}

All logarithms are base $2$, while $\ln$ denotes the natural
logarithm. We write
\[
 L_{\mathrm{in}}=\log(|\mathcal X||\mathcal Y|),\qquad
 L_{\mathrm{out}}=\log(|\mathcal A||\mathcal B|),
\]
for the logarithms of the input and output alphabet sizes. The binary
entropy is $h(t)=-t\log t-(1-t)\log(1-t)$, with
$0\log0=0$. For distributions $q,P$ on the same finite set,
\[
 D(q\Vert P)=\sum_{z:q(z)>0}q(z)\log\frac{q(z)}{P(z)},
 \qquad
 \TV(q,P)=\frac12\sum_z|q(z)-P(z)|.
\]
Set $D(q\Vert P)=+\infty$ if $q(z)>0=P(z)$ for some $z$.
A distribution written as a subscript on an entropy or information
quantity, e.g., $H_\mu(X \mid Y)$, specifies the probability experiment used. 

For input strings $X^n$, we use
$X_{<i}=(X_1,\ldots,X_{i-1})$,
$X_{>i}=(X_{i+1},\ldots,X_n)$, and
$X_{\le i}=(X_1,\ldots,X_i)$, and analogously for $Y^n$.
Also $[n]=\{1,\ldots,n\}$.

\subsection{Probability and information facts}
\label{sec:basic-facts}

We use the following standard identities and inequalities for finite
distributions. For~\cref{fact:basic-information,fact:mutual-information},
see Ref.~\cite[Chapter 2]{CT91}; for Pinsker's inequality, see
Ref.~\cite[Problem 3.18]{CK11}.

\begin{Fact}
\label{fact:basic-information}
For finite distributions $q,P$, relative entropy is nonnegative and
satisfies the chain rule
\begin{equation}
 D(q_{XY}\Vert P_{XY})
 =D(q_X\Vert P_X)
  +\E_{x\sim q_X}D(q_{Y\mid X=x}\Vert P_{Y\mid X=x}).
 \label{eq:relative-entropy-chain}
\end{equation}
If $P(E)>0$, then
$D(P(\,\cdot\mid E)\Vert P)=\log(1/P(E))$.
\end{Fact}

\begin{Fact}
\label{fact:mutual-information}
Conditional mutual information is nonnegative and satisfies
\begin{align}
 I_q(X;V\mid Y)
 &=\E_{(x,y)\sim q}D(q_{V\mid x,y}\Vert q_{V\mid y})
 =H_q(X\mid Y)-H_q(X\mid Y,V),\notag\\
 I_q(X;V,W\mid Y)
 &=I_q(X;V\mid Y)+I_q(X;W\mid Y,V).
 \label{eq:mutual-information-chain}
\end{align}
Applying the same randomized map to two distributions cannot increase
their relative entropy or total variation.
\end{Fact}

\begin{Fact}[Pinsker's inequality]
\label{fact:pinsker}
For finite distributions,
\[
 \TV(q,P)\le\sqrt{\frac{\ln2}{2}D(q\Vert P)}.
\]
For any event $F$, $|q(F)-P(F)|\le\TV(q,P)$.
\end{Fact}

The next inequality bounds the change in information cost when
two joint distributions are close. Its dependence on the input
alphabets does not involve the alphabet of the transcript. The
conditional-entropy bound follows from the stronger result in
Ref.~\cite{AS20}.
\begin{Fact}
\label{fact:continuity}
If $\TV(q_{ZW},P_{ZW})\le\delta\le1/2$, then
\begin{equation}
 |H_q(Z\mid W)-H_P(Z\mid W)|
 \le\delta\log|\mathcal Z|+h(\delta).
 \label{eq:conditional-entropy-continuity}
\end{equation}
Consequently, if $\TV(q_{XYH},P_{XYH})\le\delta\le1/2$, then
\begin{align}
 &\bigl|I_q(X;H\mid Y)+I_q(Y;H\mid X)
       -I_P(X;H\mid Y)-I_P(Y;H\mid X)\bigr|\notag\\
 &\qquad\le2L_{\mathrm{in}}\delta+4h(\delta).
 \label{eq:information-continuity}
\end{align}
\end{Fact}

The information bound~\eqref{eq:information-continuity} follows by
applying the conditional-entropy bound
\eqref{eq:conditional-entropy-continuity} to
$H(X\mid Y)$, $H(X\mid Y,H)$, $H(Y\mid X)$, and $H(Y\mid X,H)$.

\subsection{Protocol facts}

We use~\cref{prop:message-cost} to lower-bound communication cost by information cost. For the standard comparison between information and communication, see Refs.~\cite{BR11,RY20} for example. 

\begin{Prop}
\label{prop:message-cost}
For any input distribution $\nu$ and finite binary protocol $\pi$,
\[
 \IC_\nu(\pi)\le\Cost_{\mathrm{wc}}(\pi).
\]
\end{Prop}

The following fact is used to simplify the explicit exponent.
Its proof appears in Appendix~\ref{app:facts}.
\begin{Fact}
\label{fact:entropy-power}
For $0\le t\le1$, the binary entropy satisfies $h(t)\le3\sqrt t$.
\end{Fact}

\section{Main result}
\label{sec:main-result}

Our main theorem, Theorem~\ref{thm:converse}, is formally stated as follows.

\begin{Thm}[Exponential strong converse]
\label{thm:converse}
Fix $0\le R<\IC_\mu(g,0)$. Then $\err_\mu(R)>0$, and every
protocol $\Pi_n$ for $g^n$ with
$\Cost_{\mathrm{wc}}(\Pi_n)\le nR$ satisfies
\begin{equation}
	\Pr_{\mu^{\otimes n}}[\Pi_n\text{ succeeds on every coordinate}]
 \le 2^{-n\kappa(R)}.
 \label{eq:strong-converse}
\end{equation}
For $R>0$, one may take
\begin{equation}
 \kappa(R)=2^{-28}\min\left\{
 \frac{\err_\mu(R)^3}{L_{\mathrm{out}}},\,
 \frac{\err_\mu(R)^4R^2}{(1+L_{\mathrm{in}})^4}
 \right\}.
 \label{eq:explicit-positive}
\end{equation}
For $R=0$, one may take
\begin{equation}
 \kappa(0)=\frac{\eta^3}{128L_{\mathrm{out}}},
 \qquad \eta=1-\omega(g,\mu)=\err_\mu(0).
 \label{eq:zero-communication}
\end{equation}
\end{Thm}

\subsection{Proof overview}\label{sec:overview}

To prove Theorem~\ref{thm:converse}, we argue by contradiction.
Fix a protocol $\Pi_n$ as in the theorem, let $p>0$ be its
success probability under $\mu^{\otimes n}$, and set
$t=\log(1/p)/n$. If $t$ is too small, we will obtain a protocol for
one instance $(g,\mu)$ with information cost at most $R$ and
error strictly below $\err_\mu(R)$. (We also prove $\err_\mu(R) > 0$ whenever $R < IC_\mu(g, 0)$.)
This contradicts the definition of \(\err_\mu(R)\). Hence \(t\) must be bounded below, yielding an exponential upper bound on \(p\).
 The construction and the explicit lower bound on $t$
are given in \cref{sec:reduction,sec:fresh-input,sec:zero-communication,sec:proof_SC}.

Below a detailed overview is explained section by section. 
In brief, we extract a single-instance distribution
$q_{\mathrm{single}}$ from the $n$-instance protocol $\Pi_n$
and use this distribution to construct a single-instance
protocol $\pi_\mathrm{single}$ that contradicts the definition of $\err_\mu(R)$.

\subsubsection*{Conditioning and selecting a coordinate (\cref{sec:reduction})}

We first extract from the $n$-instance protocol $\Pi_n$ a joint
distribution $q$ of a single input pair $(X,Y)$, output pair $(A,B)$,
and auxiliary variable $V$. The distribution $q$ need not arise from
a one-instance protocol. Nevertheless, we define its error by
$q\{(A,B)\notin g(X,Y)\}$ and consider the expression
$I_q(X;V\mid Y)+I_q(Y;V\mid X)$, using the same formulas as for a
protocol distribution.

We need $q$ to have four properties: (P.1) its error is at most
$\varepsilon$; (P.2) its input marginal $q_{XY}$ is close to $\mu$;
(P.3) the information expression above is at most $R$ plus a vanishing term; and (P.4) its outputs $AB$ can be
sampled locally given
$V$ and the respective inputs. These properties control the error and
information cost of the protocol constructed in \cref{sec:fresh-input}
and the error of the zero-message protocol constructed in
\cref{sec:zero-communication}.  \Cref{sec:reduction} constructs $q$ and proves the required properties.
Below we explain the construction of $q$ and how those four properties are proved in detail.

\emph{Construction of $q$.} We begin with the probability distribution induced by $\Pi_n$.
Fix its private and public randomness in a way that the resulting deterministic protocol still
succeeds on every coordinate with probability at least $p$. Run this
protocol on $(X^n,Y^n)\sim\mu^{\otimes n}$, and denote its transcript and the
players' output strings by $T,A^n,B^n$. Let $P$ be the joint
distribution of these variables and the inputs. For the following explanation, for each $i\in[n]$,
let $W_i$ be the event $(A_i,B_i)\in g(X_i,Y_i)$ that $\Pi_n$
succeeds on coordinate $i$. Thus
$P(\bigcap_{i=1}^n W_i)\ge p$.

We now construct an event $W_S$ conditioned on which the protocol $\Pi_n$ succeeds on each coordinate with probability at least $1 - \varepsilon$. For this purpose,
\cref{sec:reduction} chooses a set $S\subseteq[n]$ and defines
$W_S=\bigcap_{i\in S}W_i$. The choice ensures the higher probability of success
$P(W_i\mid W_S)\ge 1-\varepsilon$ for every $i\in[n]$, as well as
\begin{equation}
 P(W_S)\ge p,
 \qquad
 |S|\le\frac{\log(1/p)}{-\log(1-\varepsilon)}.
\end{equation}
The bound on probability limits the change caused by conditioning;
the bound on size is later used to limit the entropy $H(A_S, B_S)$. 

Write $Q=P(\,\cdot\mid W_S)$ and construct the distribution $q$ as follows. Under $Q$, choose $J$ uniformly from
$[n]$, independently of all the other variables, and define
\[
 (X,Y,A,B)=(X_J,Y_J,A_J,B_J),
 \qquad
 V=(J,X_{<J},Y_{>J},T,A_S,B_S).
\]
Finally let $q$ be the joint distribution of $(X,Y,A,B,V)$, defined based on $Q$.

\emph{The properties (P.1) to (P.4).} 
We now state the four properties precisely. Properties (P.1) and (P.2) proved in~\cref{sec:reduction} are
\begin{equation}
 q\{(A,B)\notin g(X,Y)\}\le\varepsilon,
 \qquad
 D(q_{XY}\Vert\mu)\le t.
 \label{eq:overview-coordinate-error-input}
\end{equation}
The error bound follows by averaging the conditional success
probabilities $Q(W_i)$ over $i$. The divergence bound, proved in
\cref{sec:reduction}, controls the change from the input marginal
$q_{XY}$ to the original distribution $\mu$ in the subsequent
protocol constructions.

Formally the property (P.3) bounds the sum of conditional mutual informations:
\begin{equation}
 I_q(X;V\mid Y)+I_q(Y;V\mid X)
 \le R+t\left(2+
       \frac{L_{\mathrm{out}}}{-\log(1-\varepsilon)}\right).
 \label{eq:overview-coordinate-information}
\end{equation}
As shown in \cref{sec:reduction}, this bound follows from the
bound for the communication cost $|\Pi_n|$, the cost of conditioning on $W_S$,
and the entropy $H(A_S,B_S)$. It will bound the information
cost of the  protocol $\pi$ constructed in \cref{sec:fresh-input} and the
sampling error of the zero-message protocol in
\cref{sec:zero-communication}.

The property (P.4) indicates local output sampling. For every
$(x,y,v)$ with $q(x,y,v)>0$, \cref{sec:reduction} proves
\begin{equation}
 q(a,b\mid x,y,v)=q(a\mid x,v)q(b\mid y,v).
 \label{eq:overview-coordinate-outputs}
\end{equation}
The variables retained in $V$ make this factorization valid even
after conditioning on $W_S$. Consequently, Alice can sample from
$q(A\mid x,v)$ and Bob can independently sample from $q(B\mid y,v)$.

\subsubsection*{A protocol on inputs from $\mu$ (\cref{sec:fresh-input})}

We next need a protocol on an input pair drawn from $\mu$ whose
distribution is close to $q$. Conditioning on $W_S$ depends on both
players' inputs, so $q$ itself need not have a valid protocol
transcript. Overcoming this issue, \Cref{sec:fresh-input} constructs a protocol $\pi$ on a
new input pair $(x,y)\sim\mu$ by sampling each message from the
appropriate conditional distribution of $q$ in the following manner:

\begin{enumerate}
\item The parties publicly choose $J=i$ and sample $X_{<i}$ from
$q(X_{<i}\mid J=i)$, independently of $(x,y)$.
\item Alice samples $Y_{>i}$ from $q(Y_{>i}\mid X=x,X_{<i},J=i)$ and sends it to Bob.
\item They then send the messages of $T$,
followed by the auxiliary outputs $A_S,B_S$, using the conditional
distributions of $q$: each sender uses their own input and the
preceding messages.
\item Finally, they sample their outputs separately
using the distributions on the right-hand side of
\eqref{eq:overview-coordinate-outputs}.
\end{enumerate}

The central comparison proven in \cref{sec:fresh-input} is that the joint
distribution $P_\pi$ of $(X,Y,A,B,V)$ generated by $\pi$ satisfies
$D(q\Vert P_\pi)\le t$, even though the input distribution for $\pi$ is $\mu$.
When $\delta=\sqrt{(\ln 2)t/2}\le 1/2$, Pinsker's
inequality then bounds the total variation distance between $q$ and
$P_\pi$ by $\delta$. Thus the error bound for $q$ in
\eqref{eq:overview-coordinate-error-input} transfers to $\pi$,
while continuity of conditional entropy transfers the information
bound in \eqref{eq:overview-coordinate-information}. Explicitly,
\begin{equation}
 \begin{split}
 \err_\mu(\pi)&\le\varepsilon+\delta,\\
 \IC_\mu(\pi)&\le I_q(X;V\mid Y)+I_q(Y;V\mid X)
                  +2L_{\mathrm{in}}\delta+4h(\delta).
 \end{split}
 \label{eq:overview-protocol-bounds}
\end{equation}
Thus \(\pi\) has error at most \(\varepsilon+\delta\) and information cost at most \(R\) plus terms tending to zero as \(t\to0\). 
However, the definition of \(\err_\mu(R)\) strictly requires information cost at most \(R\), so this bound~\eqref{eq:overview-protocol-bounds} on $\IC_\mu(\pi)$ alone does not yield the desired contradiction.

\subsubsection*{A zero-message protocol (\cref{sec:zero-communication})}
We next construct a zero-message protocol \(\sigma\), which trivially has information cost zero. Using the zero-information protocol,  a public mixture of \(\pi\) and \(\sigma\) will reduce any excess information cost to meet the budget \(R\) for $R > 0$. For \(R=0\), we use \(\sigma\) alone.
\Cref{sec:zero-communication} constructs such a protocol $\sigma$.
The core idea of the construction is that on input $(x,y)$, Alice and Bob use shared
randomness to obtain correlated samples $V_A, V_B$ from $q(V\mid x)$ and
$q(V\mid y)$, respectively, without sending messages. Each player
then uses the sampled value and their own input to produce an output.

For the error analysis, we consider a sample $V_*$ from $q(V\mid x,y)$, which the players do not need to
compute. Hypothetical outputs using this third sample $V_*$ have the
conditional distribution $q(A,B\mid x,y)$ by
the local output condition \eqref{eq:overview-coordinate-outputs}. 
Therefore the parties output correctly following the distribution $q$ whenever $V_A = V_B = V_*$.

We then analyze the error probability of $\sigma$. \Cref{sec:zero-communication} bounds the error probability using the two conditional
mutual informations in
\eqref{eq:overview-coordinate-information}. An intermediate step gives an
error bound under inputs drawn from $q_{XY}$. To transfer this to an error bound on $\mu$, we apply the input-distribution
bound in \eqref{eq:overview-coordinate-error-input} giving
$\operatorname{TV}(q_{XY},\mu)\le\delta$. The resulting bound is
\begin{equation}
 \err_\mu(\sigma)
 \le\varepsilon+\delta+
  2\sqrt{(\ln 2)[I_q(X;V\mid Y)+I_q(Y;V\mid X)]},
 \qquad \IC_\mu(\sigma)=0.
 \label{eq:overview-zero-message}
\end{equation}

\subsubsection*{Completing the contradiction (\cref{sec:proof_SC})}

We now combine the two protocols $\pi, \sigma$ to obtain information cost at most
$R$ without losing the small-error guarantee when $t$ is small.
\Cref{sec:proof_SC} gives the combination as follows.
If $C =\IC_\mu(\pi)\le R$,
use $\pi$. If $C>R>0$, use a public coin to run $\pi$ with probability
$R/C$ and $\sigma$ otherwise. The mixed protocol has information
cost $R$. Regarding the error bound, by \eqref{eq:overview-coordinate-information} and the
information-cost inequality in \eqref{eq:overview-protocol-bounds},
$C-R$ is at most a sum of terms tending to zero with $t$. Hence
the probability $1-R/C\le(C-R)/R$ of using $\sigma$ also tends to zero as $t \to 0$.
Therefore the error of the mixed protocol approaches at most $\varepsilon$:
the error of $\pi$ is at most $\varepsilon+\delta =\varepsilon +o(1)$ as $t\to 0$, and the
probability of using $\sigma$ tends to zero. (If $R=0$, use
$\sigma$ alone and apply \eqref{eq:overview-zero-message}.)
In either case, sufficiently small $t$ would give information
cost at most $R$ and error below $\err_\mu(R)$ by choosing sufficiently small $\varepsilon$. 
\Cref{sec:proof_SC} also chooses
the parameters explicitly to obtain the exponents in
Theorem~\ref{thm:converse}. 

\subsubsection*{Consequences (\cref{sec:consequences})}

Finally, \cref{sec:consequences} combines our main result with
communication upper bounds to determine the asymptotic
distributional communication rate. It then uses the zero-error
minimax identity to derive the corresponding  strong
converse and asymptotic rate on randomized communication complexity.

\section{Conditioning and selecting a coordinate}
\label{sec:reduction}

Throughout \cref{sec:reduction}, fix a protocol $\Pi_n$ for $g^n$ that
communicates at most $nR$ bits and succeeds on $\mu^{\otimes n}$ with probability
at least $p>0$. By fixing its public and private randomness, we may
assume that $\Pi_n$ is deterministic while preserving these two bounds.

For each $0<\varepsilon<1$, we construct a distribution
$q(X,Y,A,B,V)$ by conditioning on success on the coordinates in $S$ and choosing a coordinate $J \in [n]$ uniformly, formally given in~\cref{constr:q}. We show that its error
$q\{(A,B)\notin g(X,Y)\}$ is at most $\varepsilon$ and that it
satisfies the following bounds:
\begin{align}
D(q_{XY}\Vert\mu)
    &\le \frac{\log(1/p)}{n}, \label{eq:reduction-input-goal}\\
I_q(X;V\mid Y)+I_q(Y;V\mid X)
    &\le R+
       \frac{\log(1/p)}{n}
       \left(2+\frac{L_{\mathrm{out}}}
                        {-\log(1-\varepsilon)}\right), \label{eq:reduction-information-goal}\\
q(a,b\mid x,y,v)
    &=q(a\mid x,v)q(b\mid y,v). \label{eq:reduction-output-goal}
\end{align}
The distribution $q$ need not arise from a protocol on a single
input pair: $V$ may not be a transcript generated from $(X,Y)$. 

The following construction fixes the variables used throughout this section.
\begin{construction}[Definition of $q$]
\label{constr:q}
Run the fixed deterministic protocol \(\Pi_n\) on
\((X^n,Y^n)\sim\mu^{\otimes n}\). Write \(T\) for its transcript and
\(A^n,B^n\) for its outputs, and let \(P\) be their joint
distribution with \(X^n,Y^n\). For \(i\in[n]\) and
\(S\subseteq[n]\), define
\[
    W_i=\{(A_i,B_i)\in g(X_i,Y_i)\},
    \qquad
    W_S=\bigcap_{i\in S}W_i.
\]
Next take a set \(S\subseteq[n]\) such that
\[
    |S|\le\frac{\log(1/p)}{-\log(1-\varepsilon)},
    \qquad
    P(W_S)\ge p,
    \qquad
    P(W_i\mid W_S)\ge 1-\varepsilon
    \quad(i\in[n])
\]
 Its existence is shown in~\cref{lem:selection}.
Fix such a set \(S\), put
\(\overline T=(T,A_S,B_S)\), and let
\(Q=P(\,\cdot\mid W_S)\). Choose \(J\) uniformly from
\([n]\), independently of the variables under \(Q\).
We define \(q(X,Y,A,B,V)\) as the distribution of
\begin{equation}
\begin{aligned}
    X&=X_J, & Y&=Y_J,\\
    A&=A_J, & B&=B_J,\\
    V&=(J,X_{<J},Y_{>J},\overline T).
\end{aligned}
\label{eq:coordinate-distribution}
\end{equation}
The variables on the right-hand side are sampled under \(Q\).
\end{construction}

We first justify the choice of $S$ and prove the bounds
needed under $Q$, including a factorization of probability similar to~\eqref{eq:reduction-output-goal}. We then
derive the displayed properties of $q$.
\subsection{Conditioning on successful coordinates}
\label{sec:conditioning}

\cref{lem:selection} proves bounds on the size of $S$ and the probability $\Pr(W_S)$ while ensuring high conditional success in every coordinate.

\begin{Lem}
\label{lem:selection}
Let $W_1,\ldots,W_n$ be the events in~\cref{constr:q}. For every $0<\varepsilon<1$, there is
a subset $S\subseteq[n]$ such that,
\begin{equation}\label{eq:lem:selection}
 |S|\le\frac{\log(1/p)}{-\log(1-\varepsilon)},\qquad
 \Pr(W_S)\ge p,\qquad
 \Pr(W_i\mid W_S)\ge1-\varepsilon\quad(i\in[n]).
\end{equation}
\end{Lem}
\begin{proof}
Start with $S=\varnothing$. Whenever an index $i\notin S$ has
$\Pr(W_i\mid W_S)<1-\varepsilon$, add it to $S$.
At termination, every coordinate satisfies the conditional
success bound in~\eqref{eq:lem:selection}. Throughout this
procedure, $W_S$ contains $\bigcap_iW_i$, so $\Pr(W_S)\ge p$.

For the upper bound on $|S|$, each addition multiplies $\Pr(W_S)$ by a factor less than $1-\varepsilon$:

\begin{equation*}
\Pr(W_S\cap W_i)
=\Pr(W_i\mid W_S)\Pr(W_S)<(1-\varepsilon)\Pr(W_S).
\end{equation*}
Since $W_S$ always contains $\bigcap_{i\le n}W_i$, when the procedure terminates,
we have
\[
p\le\Pr(W_S)\le(1-\varepsilon)^{|S|}.
\]
Rearranging gives the bound on $|S|$.
\end{proof}

\paragraph{The bounds on information quantities.}
For the set $S$ fixed in \cref{constr:q}, we have
$Q(W_i)=P(W_i\mid W_S)\ge1-\varepsilon$ for every $i$.
Let $\nu=Q_{X^nY^n}$. Since the protocol is deterministic,
$W_S$ and all transcript and output variables are determined by
$(X^n,Y^n)$. Thus $P$ and $Q$ have the same conditional distribution of
these variables given the inputs, and $\nu$ is $\mu^{\otimes n}$
conditioned on $W_S$. Consequently,
\begin{equation}
 D(Q\Vert P)=D(\nu\Vert\mu^{\otimes n})
 =\log\frac1{P(W_S)}
 \le\log\frac1p.
 \label{eq:conditioning-cost}
\end{equation}
The original deterministic protocol has worst-case cost at
most $nR$ even when the inputs follow $\nu$. Therefore 
Proposition~\ref{prop:message-cost} shows that its information cost under the input distribution $Q_{X^n Y^n}$ is
also at most $nR$. Appending $A_S$ and $B_S$ to $T$, yielding $\overline{T} =(A_S, B_S, T)$, contributes at most their
combined alphabet entropy, which is at most $|S|L_{\mathrm{out}}$. Hence
\begin{equation}
 I_Q(X^n;\overline T\mid Y^n)
 +I_Q(Y^n;\overline T\mid X^n)
 \le nR+|S|L_{\mathrm{out}}.
 \label{eq:augmented-information}
\end{equation}

\paragraph{Locality of outputs after conditioning.}
The event $W_S$ depends on both players' inputs, so conditioning on
it might appear to destroy the locality of their outputs; the output A may depend on Bob's input $Y$ for example. 
For coordinate $i$, reveal the selected outputs together with $X_{<i}$
and $Y_{>i}$, and put
$V_i=(X_{<i},Y_{>i},\overline T)$. Alice's remaining input is
$X_{>i}$ and Bob's is $Y_{<i}$. The next lemma shows that these
parts are conditionally independent in the form needed for the
coordinate outputs.

\begin{Lem}\label{lem:local-outputs}
For every $i\in[n]$, the distribution $Q$ of \cref{constr:q} satisfies
\begin{equation}
 Q(A_i,B_i\mid X_i,Y_i,V_i)
 =Q(A_i\mid X_i,V_i)Q(B_i\mid Y_i,V_i)
 \label{eq:coordinate-output-locality}
\end{equation}
at every conditioning value of positive $Q$-probability.
\end{Lem}
\begin{proof}
If $i\in S$, then $A_i$ and $B_i$ are already determined by
$V_i$, and the identity follows. Suppose $i\notin S$, and fix
$(X_i,Y_i,V_i)=(x,y,v_i)$ with positive $Q$-probability, where
$v_i=(x_{<i},y_{>i},\bar t)$.

The selected outputs $A_S, B_S$ in $\bar t$ can be viewed as additional
messages sent after $T$. For each fixed $\bar t$, let
$\mathcal C_A(\bar t)\subseteq\mathcal X^n$ be the Alice inputs
consistent with her messages and selected outputs, and let
$\mathcal C_B(\bar t)\subseteq\mathcal Y^n$ be the analogous Bob
inputs. The rectangle property of the deterministic transcript
gives
\begin{equation}
 P(x^n,y^n,\bar t)
 =\mu^{\otimes n}(x^n,y^n)
  \mathbf1\{x^n\in\mathcal C_A(\bar t)\}
  \mathbf1\{y^n\in\mathcal C_B(\bar t)\}.
 \label{eq:lem:local-outputs}
\end{equation}

Write $(a_s,b_s)_{s\in S}$ for the outputs recorded in $\bar t$.
For a possible suffix $x_{>i}$ and prefix $y_{<i}$, define
\begin{align*}
 F_{x,v_i}(x_{>i})
 &=\mathbf1\{(x_{<i},x,x_{>i})\in\mathcal C_A(\bar t)\}
   \prod_{j>i}\mu(x_j,y_j)
   \prod_{\substack{s\in S\\s>i}}
      \mathbf1\{(a_s,b_s)\in g(x_s,y_s)\},\\
 G_{y,v_i}(y_{<i})
 &=\mathbf1\{(y_{<i},y,y_{>i})\in\mathcal C_B(\bar t)\}
   \prod_{j<i}\mu(x_j,y_j)
   \prod_{\substack{s\in S\\s<i}}
      \mathbf1\{(a_s,b_s)\in g(x_s,y_s)\}.
\end{align*}
All $y_j$ in $F_{x,v_i}$ and all $x_j$ in $G_{y,v_i}$
are fixed by $v_i$. Because $i\notin S$, no success indicator
involves both $x$ and $y$. In particular, $F_{x,v_i}$ does not
depend on $y$, and $G_{y,v_i}$ does not depend on $x$.

The product form of $\mu^{\otimes n}$, the rectangle identity
\eqref{eq:lem:local-outputs}, and the definition of $W_S$ now give
\begin{align*}
 &P(X_{>i}=x_{>i},Y_{<i}=y_{<i},X_i=x,Y_i=y,V_i=v_i,W_S)\\
 &\qquad=\mu(x,y)F_{x,v_i}(x_{>i})G_{y,v_i}(y_{<i}).
\end{align*}
The fixed value $(x,y,v_i)$ has positive $Q$-probability, so the
two sums below are positive. Normalizing the displayed product
yields
\begin{align}
 Q(x_{>i},y_{<i}\mid x,y,v_i)
 &=
 \frac{F_{x,v_i}(x_{>i})}
      {\sum_{x'_{>i}}F_{x,v_i}(x'_{>i})}
 \frac{G_{y,v_i}(y_{<i})}
      {\sum_{y'_{<i}}G_{y,v_i}(y'_{<i})}.
 \label{eq:remaining-inputs}
\end{align}
The first factor is independent of $y$. Marginalizing over
$Y_{<i}$, then averaging over $Y_i$ conditional on $(X_i,V_i)$,
identifies that factor with $Q(x_{>i}\mid x,v_i)$. Similarly,
the second factor is $Q(y_{<i}\mid y,v_i)$. Thus
\[
 Q(x_{>i},y_{<i}\mid x,y,v_i)
 =Q(x_{>i}\mid x,v_i)Q(y_{<i}\mid y,v_i).
\]
Given $(x,y,v_i)$, Alice's output $A_i$ depends only on
$x_{>i}$, and Bob's output $B_i$ depends only on $y_{<i}$.
Applying their output functions to this product distribution
proves \eqref{eq:coordinate-output-locality}.
\end{proof}

\subsection{The conditioned coordinate distribution}

We now verify the properties of the distribution $q$ in
\cref{constr:q}. Write $\err(q)=q\{(A,B)\notin g(X,Y)\}$.

\begin{Lem}
\label{lem:coordinate-information}
The distribution $q$ of \cref{constr:q} satisfies
$\err(q)\le\varepsilon$ and
\eqref{eq:reduction-input-goal}--\eqref{eq:reduction-output-goal}.
\end{Lem}

\begin{proof}
	We first prove $\err(q)\le\varepsilon$ and \eqref{eq:reduction-output-goal}, and then establish \eqref{eq:reduction-information-goal} and \eqref{eq:reduction-input-goal}.
\proofstep{Error and output locality.}
By \cref{constr:q}, $Q(W_i)\ge1-\varepsilon$ for every $i$.
Since $J$ is uniform, the success probability under $q$ is
$n^{-1}\sum_iQ(W_i)$; hence $\err(q)\le\varepsilon$.
For each fixed $J=i$, the factorization
\eqref{eq:coordinate-output-locality} under $Q$ applies with
$V_i=(X_{<i},Y_{>i},\overline T)$.
Since \cref{constr:q} implies
\begin{equation*}
Q(A_i=a,B_i=b \mid X_i= x, Y_i =y,V_i = v)= q(a, b \mid x,y,V= (J=i,v)),
\end{equation*}
the factorization under $Q$ yields the factorization~\eqref{eq:reduction-output-goal} under $q$ as desired.

\proofstep{Information bound.}
Our intermediate goal is 
\begin{equation}
 nI_q(X;V\mid Y)\le I_Q(X^n;\overline T\mid Y^n)+D(Q_{X^nY^n}\Vert\mu^{\otimes n}).
 \label{eq:inter:lem:bound}
\end{equation}
For this purpose, we will now bound each term in $nI_q(X;V \mid Y) = nH_q(X \mid Y) - nH_q(X \mid V,Y)$ separately;
we first upper-bound $nH_q(X \mid Y)$ and then lower-bound $nH_q(X \mid Y,V)$.

Nonnegativity of conditional relative
entropy 
\begin{equation*}
	0 \le \E_{q_Y} D(q_{X\mid Y} \Vert \mu_{X \mid Y})
	= \E_q[- \log \mu(X \mid Y)] - H_q(X \mid Y)
\end{equation*}
and the equality $q_{XY} = \E_J[Q_{X_J Y_J}]$ give
\begin{align*}
 nH_q(X\mid Y)
 &\le n \cdot \E_q[- \log \mu(X \mid Y)]\\
 &= \sum_i \E_Q[-\log\mu(X_i\mid Y_i)]\\
 &=\E_Q[-\log\mu^{\otimes n}(X^n\mid Y^n)].
\end{align*}
On the other hand, $H(X \mid Y Z) \le H(X \mid Y)$ holds for any random variables $X, Y, Z$. Therefore decomposing $V= (J, X_{<i}, Y_{>i}, \overline T)$ and the chain rule conditioned on $J=i$ give
\begin{align*}
 nH_q(X\mid Y,V)
 &=\sum_i H_Q(X_i\mid X_{<i},Y_i,Y_{>i},\overline T)\\
 &\ge \sum_i H_Q(X_i\mid X_{<i},Y_{<i} Y_i,Y_{>i},\overline T)\\
 &= H_Q(X^n\mid Y^n,\overline T).
\end{align*}
Subtracting these bounds and applying the relative entropy chain
rule of~\cref{fact:basic-information} yields
\begin{align*}
 nI_q(X;V\mid Y)
 &= n\left[ H_q(X \mid Y) - H_q(X \mid V, Y)\right] \\
 &\le \E_Q[-\log\mu^{\otimes n}(X^n\mid Y^n)]
       -H_Q(X^n\mid Y^n,\overline T)\\
 &= \bigl(H_Q(X^n \mid Y^n)-H_Q(X^n\mid Y^n,\overline T) \bigr) 
 - \bigl(H_Q(X^n \mid Y^n) - \E_Q[-\log\mu^{\otimes n}(X^n\mid Y^n)]\bigr)\\
 &=I_Q(X^n;\overline T\mid Y^n)
   +D(Q_{X^nY^n}\Vert\mu^{\otimes n})
   -D(Q_{Y^n}\Vert\mu_Y^{\otimes n})\\
 &\le I_Q(X^n;\overline T\mid Y^n)+D(Q_{X^nY^n}\Vert\mu^{\otimes n})
\end{align*}
as desired in~\eqref{eq:inter:lem:bound}.

Interchanging $X$ and $Y$ similarly gives
\[
 nI_q(Y;V\mid X)
 \le I_Q(Y^n;\overline T\mid X^n)+D(Q_{X^nY^n}\Vert\mu^{\otimes n}).
\]
Therefore,
the divergence bound~\eqref{eq:conditioning-cost} and the information bound~\eqref{eq:augmented-information} on $T$ imply
\[
 n\bigl[I_q(X;V\mid Y)+I_q(Y;V\mid X)\bigr]
 \le nR+|S|L_{\mathrm{out}}+2\log\frac1p.
\]
The bound on $|S|$ in \cref{constr:q} proves
the desired information bound \eqref{eq:reduction-information-goal}.

\proofstep{Input marginal.}
Apply the relative entropy chain rule of
\cref{fact:basic-information} successively to the coordinate
pairs $(X_i,Y_i)$. Since the joint distribution $\mu^{\otimes n}$ is a
product, the conditional distribution $\mu(X_i Y_i \mid X_{<i}Y_{<i})$ at each step is $\mu$.
Thus
\[
 D(Q_{X^nY^n}\Vert\mu^{\otimes n})
 =\sum_{i=1}^n
   \E_{Q_{X_{<i}Y_{<i}}}
   D(Q_{X_iY_i\mid X_{<i}Y_{<i}}\Vert\mu).
\]
For $i=1$, the prefix is empty and the summand is
$D(Q_{X_1Y_1}\Vert\mu)$. For each $i$, splitting the logarithm
inside the conditional divergence gives
\begin{align*}
 &\E_{Q_{X_{<i}Y_{<i}}}
   D(Q_{X_iY_i\mid X_{<i}Y_{<i}}\Vert\mu)\\
 &\quad=
 \E_Q\left[
   \log\frac{Q(X_i,Y_i\mid X_{<i},Y_{<i})}
                 {Q_{X_iY_i}(X_i,Y_i)}
 \right]
 +\E_Q\left[
   \log\frac{Q_{X_iY_i}(X_i,Y_i)}
                 {\mu(X_i,Y_i)}
 \right]\\
 &\quad=
 I_Q\bigl((X_i,Y_i);(X_{<i},Y_{<i})\bigr)
 +D(Q_{X_iY_i}\Vert\mu)\\
 &\quad \ge
 D(Q_{X_iY_i}\Vert\mu)\\
\end{align*}
 Finally,
$q_{XY}=n^{-1}\sum_iQ_{X_iY_i}$, and convexity of relative
entropy gives
\[
 D(q_{XY}\Vert\mu)
 \le\frac1n\sum_iD(Q_{X_iY_i}\Vert\mu)
 \le\frac1nD(Q_{X^nY^n}\Vert\mu^{\otimes n})
 \le\frac{\log(1/p)}{n}.
\]
This proves \eqref{eq:reduction-input-goal}.
\end{proof}

\section{A protocol under the original input distribution}
\label{sec:fresh-input}

\cref{sec:fresh-input} constructs a protocol $\pi$ which works under the distribution $\mu$, based on $q$.
The key step is to bound the divergence between its joint distribution
and the conditioned coordinate distribution $q$ of~\cref{constr:q}.
\begin{construction}[The single-instance protocol $\pi$]
\label{constr:pi}
Let $q$ be the distribution in \cref{constr:q}, with
$V=(J,X_{<J},Y_{>J},\overline T)$. 
Represent the augmented transcript $\overline T$ as alternating binary messages
$M^1,\ldots,M^m$, with Alice sending the odd messages.
(Insert fixed empty messages to handle consecutive turns by one party and terminated paths.) On inputs $(x,y)\sim\mu$,
the protocol $\pi$ proceeds as follows:
\begin{enumerate}
\item Publicly choose $J=i$ uniformly in $[n]$, and sample
      $X_{<i}$ according to $q(X_{<i}\mid J=i)$, independently
      of $(x,y)$.
\item Alice samples $Y_{>i}$ according to
      $q(Y_{>i}\mid X=x,X_{<i},J=i)$ and sends it to Bob.
\item For each $t$, Alice samples an odd message and Bob samples
      an even message using, respectively,
      \[
       \begin{aligned}
       &q(M^t\mid X=x,J=i,X_{<i},Y_{>i},M^{<t}),\\
       &q(M^t\mid Y=y,J=i,X_{<i},Y_{>i},M^{<t}).
       \end{aligned}
      \]
\item With $V=(i,X_{<i},Y_{>i},M^1,\ldots,M^m)$,
      Alice samples $A$ from $q(A\mid x,V)$ and Bob samples
      $B$ from $q(B\mid y,V)$ using independent private
      randomness.
\end{enumerate}
At values of conditional probability zero, define the message distributions arbitrarily.
Write $P_\pi$ for the joint distribution of
$(X,Y,A,B,V)$ under $\pi$.
\end{construction}

For $n\ge1$ and $0<p\le1$, define
\begin{equation}
 \delta(n,p):=\sqrt{\frac{\ln2}{2n}\log\frac1p}.
 \label{eq:input-transfer-delta}
\end{equation}

\begin{Lem}
\label{lem:communicating-coordinate}
Suppose $q$ is defined as in \cref{constr:q} and
$\delta(n,p)\le\frac12$. The protocol $\pi$ of
\cref{constr:pi} satisfies
\begin{align}
 \err_\mu(\pi)
 &\le\err(q)+\delta(n,p),
 \label{eq:communicating-error}\\
 \IC_\mu(\pi)
 &\le I_q(X;V\mid Y)+I_q(Y;V\mid X)\notag\\
 &\quad+2L_{\mathrm{in}}\delta(n,p)+4h(\delta(n,p)).
 \label{eq:communicating-information}
\end{align}
\end{Lem}

\begin{proof}
Our primary goal is to prove
\begin{equation*}
 D(q\Vert P_\pi)\le\frac{\log(1/p)}n.
\end{equation*}
The error and information bounds will then follow from
Pinsker's inequality and the information-continuity bound.
The main step is to compare the distributions $q, P_\pi$ conditional
on each choice of $J$ and bound the sum of their divergences.

Recall that $Q=P(\,\cdot\mid W_S)$ from \cref{constr:q}.
For each $i$, let $q_i$ be the distribution of
$(X_{\le i},Y_{\ge i},\overline T)$ under $Q$.
This is also the distribution of the corresponding
variables under $q$ conditional on $J=i$, with the outputs
omitted.
Let $P_\pi^{(i)}$ be their distribution under $\pi$
conditional on $J=i$, before the outputs are sampled.
We use $P$ and $Q$ also for their marginals on
$(X^n,Y^n,\overline T)$.
We will later use the fact that omitting
the other variables, i.e., the outputs, does not change $D(Q\Vert P)$,
because the original protocol is deterministic.

To reduce the desired bound to these coordinate
distributions, note that $J$ is uniform under both
$q$ and $P_\pi$.
Moreover, by the local output condition \eqref{eq:reduction-output-goal}, the
conditional output distributions agree.
The relative-entropy chain rule therefore gives
\begin{equation}
	\label{eq:lem:aver:id}
 D(q\Vert P_\pi)
 =\frac1n\sum_{i=1}^n D(q_i\Vert P_\pi^{(i)}).
\end{equation}
It thus suffices to prove
\[
 \sum_{i=1}^n D(q_i\Vert P_\pi^{(i)})
 \le D(Q\Vert P)
 \le \log(1/p),
\]
where the second inequality is already given in~\eqref{eq:conditioning-cost}.

To prove the summed divergence bound above, we will construct a
probability distribution $Q_n$ on
$(X^n,Y^n,\overline T)$ satisfying
\begin{equation}
 \frac{Q_n(x^n,y^n,m)}{P(x^n,y^n,m)}
 =
 \prod_{i=1}^n
 \frac{q_i(x_{\le i},y_{\ge i},m)}
      {P_\pi^{(i)}(x_{\le i},y_{\ge i},m)}
 \label{eq:identity:Qn}
\end{equation}
on the support of $Q$.
Because $q_i$ is the corresponding marginal of $Q$,
taking logarithms and expectations under $Q$ will then give
\begin{equation*}
 \sum_{i=1}^n D(q_i\Vert P_\pi^{(i)})
 = \mathbb E_Q\log\frac{Q_n}{P}
 = D(Q\Vert P)-D(Q\Vert Q_n)
 \le D(Q\Vert P)
\end{equation*}
as desired.

\proofstep{Construction of $Q_n$.}
We construct $Q_n$ by successive sampling.
First sample $(X^n,Y_n,\overline T)$ from $q_n$.
Then, for $i=n-1,\ldots,1$, sample $Y_i$ according to
$Q(Y_i\mid X_{\le i},Y_{>i},\overline T)$.
Conditional distributions at conditioning values of
zero $Q$-probability may be chosen arbitrarily.
This procedure defines a probability distribution
\[
 Q_n(x^n,y^n,m)
 =
 q_n(x^n,y_n,m)
 \prod_{i=1}^{n-1}
 Q(y_i\mid x_{\le i},y_{>i},m).
\]
For every $(x^n,y^n,m)$ in the support of $Q$, all factors
in this expression are positive.
Moreover, since
$q_i(x_{\le i},y_{\ge i},m)
 =Q(x_{\le i},y_{\ge i},m)$,
rewriting each conditional probability as a ratio gives
\begin{align*}
 Q_n(x^n,y^n,m)
 &=
 q_n(x^n,y_n,m)
 \prod_{i=1}^{n-1}
 \frac{Q(x_{\le i},y_{\ge i},m)}
      {Q(x_{\le i},y_{>i},m)}\\
 &=
 \frac{\prod_{i=1}^n q_i(x_{\le i},y_{\ge i},m)}
      {\prod_{i=1}^{n-1}Q(x_{\le i},y_{>i},m)}.
\end{align*}
The denominator contains the marginals shared by
consecutive coordinate distributions.
We next simplify $\prod_iP_\pi^{(i)}$ and show that
the same marginal factors occur in its expression,
which will establish \eqref{eq:identity:Qn}.

\proofstep{Product identity.}
Our goal in this step is to simplify the product $P_\pi^{(i)}$ of probability distributions.
More precisely, we will show that, on the support of $Q$,
\begin{equation*}
 \prod_{i=1}^n P_\pi^{(i)}(x_{\le i},y_{\ge i},m)
 =
 P(x^n,y^n,m)
 \prod_{i=1}^{n-1}Q(x_{\le i},y_{>i},m).
\end{equation*}
Comparing this expression with the product expression
for $Q_n$ above will prove \eqref{eq:identity:Qn}.

Fix $(x^n,y^n,m)$ with $Q(x^n,y^n,m)>0$.
We use the conventions
\[
 Q(x_{<1})=Q(y_{>n})=1,
 \qquad
 Q(y_{>n}\mid x^n)=1,
 \qquad
 Q(x_{\le n},y_{>n})=Q(x^n).
\]
The definition of $q$ in \cref{constr:q} and the sampling
steps of $\pi$ in \cref{constr:pi} give
\begin{align}
 &P_\pi^{(i)}(x_{\le i},y_{\ge i},m)\notag\\
 &\quad=\mu(x_i,y_i)Q(x_{<i})Q(y_{>i}\mid x_{\le i})\notag\\
 &\qquad\quad\cdot
   \prod_{t\ \mathrm{odd}}
      Q(m^t\mid x_{\le i},y_{>i},m^{<t})
   \prod_{t\ \mathrm{even}}
      Q(m^t\mid x_{<i},y_{\ge i},m^{<t}).
 \label{eq:coordinate-probability}
\end{align}
Every factor is positive under the stated assumption.
In the following, we evaluate the product of the input factors and the product
of the message factors in \eqref{eq:coordinate-probability}
separately.

We begin with the input factors involving
$X_{<i}$ and $Y_{>i}$:
\begin{align*}
 \prod_{i=1}^n Q(x_{<i})Q(y_{>i}\mid x_{\le i})
 &=
 \left(
   \prod_{i=1}^n\frac{Q(x_{<i})}{Q(x_{\le i})}
 \right)
 \prod_{i=1}^n Q(x_{\le i},y_{>i})\\
 &=
 \frac{1}{Q(x^n)}
 \prod_{i=1}^n Q(x_{\le i},y_{>i})\\
 &=\prod_{i=1}^{n-1}Q(x_{\le i},y_{>i}).
\end{align*}
In the second equality, the numerator at index $i+1$
cancels the denominator at index $i$, leaving $1/Q(x^n)$.
The last equality cancels this remaining denominator
against $Q(x_{\le n},y_{>n})=Q(x^n)$.
These equalities also hold for $n=1$ under the above
conventions, with the final empty product equal to one.

For the factors including $m$'s, pair the odd-message factors
at coordinate $i$ with the even-message factors at
coordinate $i+1$. Since $x_{<i+1}=x_{\le i}$ and
$y_{\ge i+1}=y_{>i}$, for each $i<n$ we obtain
\begin{align*}
 &\prod_{t\ \mathrm{odd}}
   Q(m^t\mid x_{\le i},y_{>i},m^{<t})
 \prod_{t\ \mathrm{even}}
   Q(m^t\mid x_{<i+1},y_{\ge i+1},m^{<t})\\
 &\qquad=
 \prod_t Q(m^t\mid x_{\le i},y_{>i},m^{<t})
 =Q(m\mid x_{\le i},y_{>i}).
\end{align*}
The remaining factors are
\[
 \prod_{t\ \mathrm{even}}
 Q(m^t\mid y^n,m^{<t})
 =
 \prod_{t\ \mathrm{odd}}
 Q(m^t\mid x^n,m^{<t})
 =1.
\]
Indeed, each such message is determined by its sender's
full input and the preceding messages, and $m$ occurs
with positive probability under $Q$.

Combining the input and message factors gives
\begin{align}
 \prod_i P_\pi^{(i)}(x_{\le i},y_{\ge i},m)
 &=\mu^{\otimes n}(x^n,y^n)
   \prod_{i=1}^{n-1}
   Q(x_{\le i},y_{>i})
   Q(m\mid x_{\le i},y_{>i})\notag\\
 &=P(x^n,y^n,m)
   \prod_{i=1}^{n-1}
   Q(x_{\le i},y_{>i},m)
 \label{eq:product-probability}
\end{align}
as desired.
Note that $P(x^n,y^n,m)=\mu^{\otimes n}(x^n,y^n)$ because the
original protocol is deterministic and
$Q(x^n,y^n,m)>0$.
Comparing \eqref{eq:product-probability} with the product
expression for $Q_n$ proves \eqref{eq:identity:Qn}.

\proofstep{Divergence bound.}
Since \eqref{eq:identity:Qn} is now established, the discussion given after~\eqref{eq:identity:Qn} gives
$\sum_iD(q_i\Vert P_\pi^{(i)})\le D(Q\Vert P)$.
The averaging identity~\eqref{eq:lem:aver:id} and
\eqref{eq:conditioning-cost} therefore imply
\begin{equation}
 D(q\Vert P_\pi)
 \le\frac{D(Q\Vert P)}n
 \le\frac{\log(1/p)}n.
 \label{eq:reduction-divergence}
\end{equation}

\proofstep{Error and information bounds.}
Pinsker's inequality (Fact~\ref{fact:pinsker}),
the divergence bound~\eqref{eq:reduction-divergence},
and the definition~\eqref{eq:input-transfer-delta} of
$\delta(n,p)$ imply
\begin{equation}
 \TV(q,P_\pi)\le\delta(n,p).
 \label{eq:reduction-distance}
\end{equation}
The probabilities of the failure event
$\{(A,B)\notin g(X,Y)\}$ under $q$ and $P_\pi$ differ
by at most this distance, proving
\eqref{eq:communicating-error}.

Under $P_\pi$, the public choice $(J,X_{<J})$ is
independent of $(X,Y)\sim\mu$, and all other components
of $V$ are messages.
Thus the definition~\eqref{eq:information-cost} gives
\[
 \IC_\mu(\pi)
 =I_{P_\pi}(X;V\mid Y)+I_{P_\pi}(Y;V\mid X).
\]
Finally, the distributions of $(X,Y,V)$ under $q$ and
$P_\pi$ have total variation distance at most
$\delta(n,p)$ by~\eqref{eq:reduction-distance}.
Applying the inequality~\eqref{eq:information-continuity} of
Fact~\ref{fact:continuity} proves
\eqref{eq:communicating-information}.
\end{proof}

\section{A zero-communication protocol}
\label{sec:zero-communication}
The protocol $\pi$ of~\cref{constr:pi} may have information cost
greater than $R$. To fix this, we construct a zero-message protocol $\sigma$ on
inputs drawn from $\mu$ and bound its error using the distribution
$q$ of~\cref{constr:q}.

\subsection{Sampling without communication}

\cref{lem:conditional-sampling} applies when, given the same value of $V$, Alice can
sample $A$ using $(X,V)$ and Bob can independently sample $B$ using
$(Y,V)$. Its proof uses the shared sampling result of~\cref{lem:shared-sampling} to generate the parties' samples of $V$ without communication.
\begin{Lem}
\label{lem:conditional-sampling}
Let $\lambda$ be a distribution of $(X,Y,V,A,B)$ with
\[
 \lambda(a,b\mid x,y,v)
 =\lambda(a\mid x,v)\lambda(b\mid y,v)
\]
whenever $\lambda(x,y,v)>0$. On inputs drawn from
$\lambda_{XY}$, there is a zero-message protocol
with error at most
\begin{equation}
 \err(\lambda)+
 2\sqrt{(\ln2)
   [I_\lambda(X;V\mid Y)+I_\lambda(Y;V\mid X)]},
 \label{eq:conditional-sampling-error}
\end{equation}
where $\err(\lambda)=\lambda[(A,B)\notin g(X,Y)]$.
\end{Lem}

\begin{proof}
	We fist construct the sampling protocol and then bound its error.
	\proofstep{Construction of the sampling protocol.}
Let $\mathcal D$ be a set of distributions:
\begin{equation*}
	\mathcal D
	= \{\lambda(V \mid x,y), \lambda(V \mid x), \lambda(V \mid y)\}_{(x,y) \in \supp(\lambda_{XY})}.
\end{equation*}
Applying Lemma~\ref{lem:shared-sampling} to the set $\mathcal D$,
we obtain a (finite-valued) shared randomness $R$ and a function $S_D: R \to \mathrm{Im} V$ such that 

\begin{equation}
	\Pr[S_{D_1}(R) \neq S_{D_2}(R)] \le 2 \TV(D_1, D_2) \qquad
	\text{for every $D_1, D_2\in\mathcal D$.}
	\label{eq:lem:conditional-sampling}
\end{equation}

On input $(x,y)$, Alice and Bob respectively set
\[
 V_A=S_{\lambda(V\mid x)}(R),
 \qquad
 V_B=S_{\lambda(V\mid y)}(R).
\]
Alice outputs according to $\lambda(A\mid x,V_A)$,
and Bob independently outputs according to
$\lambda(B\mid y,V_B)$. They send no messages.
For the analysis, define the random variable $V_*$ as
\[
 V_*=S_{\lambda(V\mid x,y)}(R).
\]
Neither player needs to compute $V_*$. When the parties successfully share the same $V_{*}$ and output $A \sim \lambda(A|V_{*},x)$ and $B \sim \lambda(B|V_{*},y)$, the error probability is $\err(\lambda)$.

\proofstep{Error bound.}
All probabilities below use the inputs $XY$ drawn from
$\lambda_{XY}$.

The bound \eqref{eq:lem:conditional-sampling} on $\TV$, Pinsker's inequality
in Fact~\ref{fact:pinsker}, and Jensen's inequality give
\begin{align*}
 \Pr[V_A\ne V_*]
 &\le 2\E_{\lambda_{XY}}
   \TV(\lambda_{V\mid X,Y},\lambda_{V\mid X})\\
 &\le \sqrt{2(\ln2)\,
   \E_{\lambda_{XY}}D(
     \lambda_{V\mid X,Y}\Vert\lambda_{V\mid X})}\\
 &=\sqrt{2(\ln2)I_\lambda(Y;V\mid X)}.
\end{align*}
Similarly,
\[
 \Pr[V_B\ne V_*]
 \le\sqrt{2(\ln2)I_\lambda(X;V\mid Y)}.
\]
A union bound then gives, 
\begin{align*}
 &\Pr[(A,B)\notin g(X,Y)]\\
 &\le\Pr\left[\{\text{$\lambda$ fails}\} \cup \{\text{Alice fails to produce $V_*$}\} \cup\{ \text{Bob fails to produce $V_*$}\}\right]\\
 &\le \err(\lambda)
   +\Pr[V_A\ne V_*]+\Pr[V_B\ne V_*],
\end{align*}
Using $\sqrt u+\sqrt v\le\sqrt{2(u+v)}$ proves
the desired inequality \eqref{eq:conditional-sampling-error}.
\end{proof}

\subsection{Running the sampler on original inputs}
We now apply \cref{lem:conditional-sampling} with $\lambda=q$ in~\cref{lem:silent-coordinate}.
Recall that the distribution $q$ of~\cref{constr:q} has
$\err(q)\le\varepsilon$ and satisfies the following properties
by~\cref{lem:coordinate-information}:
\begin{align*}
D(q_{XY}\Vert\mu)
    &\le \frac{\log(1/p)}{n}, \\
I_q(X;V\mid Y)+I_q(Y;V\mid X)
    &\le R+
       \frac{\log(1/p)}{n}
       \left(2+\frac{L_{\mathrm{out}}}
                        {-\log(1-\varepsilon)}\right), \\
q(a,b\mid x,y,v)
    &=q(a\mid x,v)q(b\mid y,v). 
\end{align*}

\begin{Lem}
\label{lem:silent-coordinate}
There is a zero-message protocol $\sigma$ on
inputs drawn from $\mu$ such that
$\IC_\mu(\sigma)=0$ and
\begin{equation}
 \begin{split}
 \err_\mu(\sigma)\le{}&
 \varepsilon+\delta(n,p)\\
 &+2\sqrt{(\ln2)\left[
   R+\frac{\log(1/p)}n \left(2+
     \frac{L_{\mathrm{out}}}{-\log(1-\varepsilon)}
       \right)\right]}.
 \end{split}
 \label{eq:silent-coordinate-error}
\end{equation}
\end{Lem}

\begin{proof}
For any input $(x,y) \in \supp(\mu)$, Alice and Bob use the
protocol from Lemma~\ref{lem:conditional-sampling}
to draw $V_A\sim q(V\mid x)$ and
$V_B\sim q(V\mid y)$. They output according to
$q(A\mid x,V_A)$ and $q(B\mid y,V_B)$,
respectively, without communicating. For inputs having zero-probability under $q$, they sample arbitrarily.
Denote this protocol by $\sigma$.

Under inputs distributed as $q_{XY}$, the consequence of~\cref{lem:conditional-sampling} is
\begin{align*}
 \operatorname{err}_{q_{XY}}(\sigma)
 &\le \err(q)+
 2\sqrt{(\ln2)
   [I_q(X;V\mid Y)+I_q(Y;V\mid X)]}\\
&\le \varepsilon 
 +2\sqrt{(\ln2)\left[
   R+\frac{\log(1/p)}n \left(2+
     \frac{L_{\mathrm{out}}}{-\log(1-\varepsilon)}
       \right)\right]},
\end{align*}
together with the properties of $q$.

To convert $\operatorname{err}_{q_{XY}}(\sigma)$ to $\operatorname{err}_{\mu}(\sigma)$, use $\TV(q_{XY}, \mu) \le \delta(n,p)$ by Pinsker's inequality. Formally, 
\begin{align*}
	|\err_{q_{XY}}(\sigma) - \err_{\mu}(\sigma)| 
	&=\left| \sum_{x,y} (q_{XY}(x,y) - \mu(x,y)) \left(\Pr[\sigma ~\text{errs} \mid x,y] - \frac12\right)\right|\\
	&\le \frac12 \sum_{x,y}|q_{XY}(x,y) - \mu(x,y)|\\
	&= \TV(q_{XY}, \mu) \le \delta(n,p).
\end{align*}
Combining the arguments proves the statement.
\end{proof}

\section{Proof of the strong converse}
\label{sec:proof_SC}

The protocol $\pi$ from~\cref{lem:communicating-coordinate}
has small error, but its information cost may exceed $R$, while the zero-message protocol $\sigma$ has larger error but its information cost is zero.
We combine the two protocols in~\cref{thm:comparison} by probabilistic mixing to obtain information cost
at most $R$. The resulting error bound will prove \cref{thm:converse}.

\subsection{A public coin protocol with information cost at most $R$}
\begin{Prop}
\label{thm:comparison}
Suppose $\Cost_{\mathrm{wc}}(\Pi_n)\le nR$, its success
probability on $\mu^{\otimes n}$ is $p>0$, and $0<\varepsilon<1$.
With $\delta(n,p)$ defined in~\eqref{eq:input-transfer-delta},
suppose $\delta(n,p)\le\frac12$.
Let $\pi$ be the protocol constructed in
Lemma~\ref{lem:communicating-coordinate}, and put
$C=\IC_\mu(\pi)$. Then the information cost $C$ satisfies the following bound:
\begin{equation}
 \begin{split}
 C\le{}& R+
 \left(2+\frac{L_{\mathrm{out}}}
                 {-\log(1-\varepsilon)}\right)
       \frac{\log(1/p)}n\\
 &+2L_{\mathrm{in}}\delta(n,p)
 +4h(\delta(n,p)).
 \end{split}
 \label{eq:comparison-cost}
\end{equation}

The error bounds on $\err_\mu(R)$ are given as follows:
If $C\le R$, then
\begin{equation}
 \err_\mu(R)\le
 \varepsilon+\delta(n,p).
 \label{eq:comparison-full}
\end{equation}
If $C>R$, then
\begin{equation}
 \begin{split}
 \err_\mu(R)\le{}&
 \varepsilon+\delta(n,p)\\
 &+2\left(1-\frac RC\right)
 \sqrt{(\ln2)\left[
 R+\left(2+
   \frac{L_{\mathrm{out}}}{-\log(1-\varepsilon)}
 \right)\frac{\log(1/p)}n\right]}.
 \end{split}
 \label{eq:comparison-mixed}
\end{equation}
\end{Prop}

\begin{proof}
	We first use the bounds established in~\cref{sec:fresh-input} and~\cref{sec:zero-communication} to obtain the bound~\eqref{eq:comparison-cost} on $C$.
	The bound~\eqref{eq:communicating-information} on information
for $\pi$ gives
\[
 C\le I_q(X;V\mid Y)+I_q(Y;V\mid X)
      +2L_{\mathrm{in}}\delta(n,p)+4h(\delta(n,p)),
\]
and the bound~\eqref{eq:reduction-information-goal} for $q$
gives
\[
 I_q(X;V\mid Y)+I_q(Y;V\mid X)
 \le R+
 \left(2+\frac{L_{\mathrm{out}}}
                 {-\log(1-\varepsilon)}\right)
       \frac{\log(1/p)}n.
\]
Substituting the second bound into the first proves
\eqref{eq:comparison-cost}.
	
For the error bounds~\eqref{eq:comparison-full}  and \eqref{eq:comparison-mixed},
recall that the error bounds for $\pi$ and $\sigma$ are
\begin{align*}
 \err_\mu(\pi)&\le\varepsilon+\delta(n,p),\\
 \err_\mu(\sigma)&\le\varepsilon+\delta(n,p)
 +2\sqrt{(\ln2)\left[
 R+\left(2+
   \frac{L_{\mathrm{out}}}{-\log(1-\varepsilon)}
 \right)\frac{\log(1/p)}n\right]}.
\end{align*}
The first uses $\err(q)\le\varepsilon$ in the error
bound~\eqref{eq:communicating-error} for $\pi$; the second
is the error bound~\eqref{eq:silent-coordinate-error} for
$\sigma$.

If $C\le R$, run $\pi$. Its information cost meets
the requirement, proving~\eqref{eq:comparison-full}.
If $C>R$, independently of the inputs, use a public coin
to run $\pi$ with probability $R/C$ and $\sigma$ otherwise.
The coin is part of the public transcript. Since
$\IC_\mu(\sigma)=0$, the information cost is
$(R/C)C=R$. Averaging the displayed error bounds gives
\eqref{eq:comparison-mixed}.
\end{proof}

\subsection{Positivity of $\err_\mu(R)$}
\cref{prop:error-curve} proves $\err_\mu(R)>0$ when $ R<\IC_\mu(g,0)$, which will be used in the proof of~\cref{thm:converse}.
At $R=0$, we also identify this error with
$\eta=1-\omega(g,\mu)$, the minimum error of a
zero-message strategy.

\begin{Prop}
\label{prop:error-curve}
If $R<\IC_\mu(g,0)$, then $\err_\mu(R)>0$.
Moreover, $\err_\mu(0)=\eta$.
\end{Prop}

\begin{proof}
We prove the first statement by contraposition. If $\err_\mu(R)=0$, the definition~\eqref{eq:def_IC_error} of $\err_\mu(R)$
supplies a sequence of protocols of information cost at most $R$
whose errors tend to zero. Thus
$\IC_\mu(g,\varepsilon)\le R$ for every
$\varepsilon>0$. The zero-error continuity in
Proposition~\ref{prop:zero-error-continuity}
implies $\IC_\mu(g,0)\le R$.

For the second statement, recall the definition of $\eta$, the minimum error probability of the game $(g,\mu)$. 
Therefore there is 
a zero-message strategy of error $\eta$, which gives
$\err_\mu(0)\le\eta$. 
It remains to show $\err_\mu(0)\ge\eta$. 
Let $\rho$
be any protocol with $\IC_\mu(\rho)=0$.
A key issue is that $\rho$ may send a lot of messages while keeping the information cost equal to zero.
To resolve this issue, we apply Lemma~\ref{lem:conditional-sampling}
to its complete distribution, taking $V$ to be
its public randomness and complete transcript. This conversion makes its communication zero.
For conditioning events of positive probability,
the output probability satisfies the factorization
\[
 \Pr[A=a,B=b\mid X=x,Y=y,V=v]
 =\Pr[A=a\mid X=x,V=v]\Pr[B=b\mid Y=y,V=v],
\]
as detailed in~Proposition~\ref{prop:protocol-locality}.
Since the public randomness is independent of the inputs,
 the sampling error overhead
 \eqref{eq:conditional-sampling-error} of~\cref{lem:conditional-sampling} is
\[
 I(X;V\mid Y)+I(Y;V\mid X)=\IC_\mu(\rho)=0.
\]
Lemma~\ref{lem:conditional-sampling} therefore gives
a zero-message strategy whose error is at most
$\err_\mu(\rho)$. It implies
$\eta\le\err_\mu(\rho)$ because $\eta$ is the minimum error probability over all zero-message strategies. This holds for any $\rho$ with zero information cost and therefore $\err_\mu(0) \ge \eta$ holds as well.
\end{proof}

\subsection{The explicit exponents}
Set $t=\log(1/p)/n$ for a protocol with success probability $p>0$.
We use~\cref{thm:comparison} and the positivity of
$\err_\mu(R)$ from~\cref{prop:error-curve} to derive the
lower bounds on $t$ stated in~\cref{thm:converse}.

\begin{proof}[Proof of Theorem~\ref{thm:converse}]
Let $\Pi_n$ be a protocol for $g^n$ with
$\Cost_{\mathrm{wc}}(\Pi_n)\le nR$ and success probability
$p$ on $\mu^{\otimes n}$. If $p=0$, the claim holds. Assume $p>0$
and write
\[
 t=\frac{\log(1/p)}n.
\]
We will bound $t$ separately for $R>0$ and $R=0$.
Since $R<\IC_\mu(g,0)$, Proposition~\ref{prop:error-curve}
gives
\[
 0<\err_\mu(R)\le\eta<1,
 \qquad \err_\mu(0)=\eta.
\]
We also have $L_{\mathrm{out}} = \log(|\mathcal{A}||\mathcal{B}|)\ge1$. Indeed, if the total
relation had only one output pair, always producing that pair
would give a zero-error protocol without messages.

\proofstep{The case $R>0$.}
We assume that $t$ is smaller than the claimed bound and
show that \cref{thm:comparison} then applies. Set
$\varepsilon=\err_\mu(R)/4$ and suppose, for a contradiction,
that
\begin{equation}
 t<2^{-28}\min\left\{
 \frac{\err_\mu(R)^3}{L_{\mathrm{out}}},
 \frac{\err_\mu(R)^4R^2}{(1+L_{\mathrm{in}})^4}
 \right\}.
 \label{eq:positive-contradiction}
\end{equation}
Because $\err_\mu(R)\le1$ and $L_{\mathrm{out}}\ge1$,
the first term in the minimum is at most one. Hence
$\delta(n,p)\le\sqrt t<2^{-14}<1/2$, as required by
\cref{thm:comparison}.

Applying~\cref{thm:comparison} provides a protocol $\pi$ whose error and
information cost are close to $\varepsilon$ and $R$,
respectively. Let $C=\IC_\mu(\pi)$, and denote the two
terms added to $R$ in the information-cost bound by
\[
 a=\left(2+\frac{L_{\mathrm{out}}}
                  {-\log(1-\varepsilon)}\right)t,
 \qquad
 b=2L_{\mathrm{in}}\delta(n,p)+4h(\delta(n,p))
\]
to simplify the expressions below.

Next we apply some elementary calculus to obtain a bound on $\err_\mu(R)$, the error achievable at
information cost at most $R$. If $C>R$, 
the inequality~\eqref{eq:comparison-mixed} gives
\[
 \err_\mu(R)\le\varepsilon+\delta(n,p)
 +2\sqrt{\ln2}\left(1-\frac RC\right)\sqrt{R+a}.
\]
Since $C\le R+a+b$ and $a,b\ge0$,
\begin{equation}
 \begin{aligned}
 \left(1-\frac RC\right)\sqrt{R+a}
 &\le\frac{(a+b)\sqrt{R+a}}{R+a+b}\\
 &\le\frac{a+b}{\sqrt{R+a}}
 \le\sqrt a+\frac b{\sqrt R}.
 \end{aligned}
 \label{eq:comparison-algebra}
\end{equation}
The last inequality uses $R+a\ge a$ and $R+a\ge R>0$.
If $C\le R$, the protocol $\pi$ itself has information
cost at most $R$ and error at most $\varepsilon+\delta(n,p)$.
Thus, in either case,
\begin{equation}
 \err_\mu(R)\le\varepsilon+\delta(n,p)
 +2\sqrt{\ln2}\left(\sqrt a+\frac b{\sqrt R}\right).
 \label{eq:positive-comparison}
\end{equation}

We now bound $a$ and $b$ in terms of $t$.
From $-\log(1-\varepsilon)\ge\varepsilon/\ln2$ and
$\varepsilon=\err_\mu(R)/4$, we get
\begin{equation}
 \begin{aligned}
 a
 &\le\left(2+\frac{4(\ln2)L_{\mathrm{out}}}
                         {\err_\mu(R)}\right)t\\
 &\le\frac{5L_{\mathrm{out}}}{\err_\mu(R)}t.
 \end{aligned}
 \label{eq:selection-polynomial}
\end{equation}
The second inequality uses $\err_\mu(R)\le1$ and
$L_{\mathrm{out}}\ge1$. Also, the entropy bound
$h(u)\le3\sqrt u$ in Fact~\ref{fact:entropy-power},
together with $\delta(n,p)\le\sqrt t$ and $t<1$, gives
\begin{equation}
 \begin{aligned}
 b
 &\le2L_{\mathrm{in}}t^{1/2}+12t^{1/4}\\
 &\le12(1+L_{\mathrm{in}})t^{1/4}.
 \end{aligned}
 \label{eq:continuity-polynomial}
\end{equation}
As $L_{\mathrm{out}}/\err_\mu(R)\ge1$, the bound
$\delta(n,p)\le\sqrt t$ is at most
$\sqrt{L_{\mathrm{out}}t/\err_\mu(R)}$.
Using the bounds on $a$ and $b$ in
\eqref{eq:positive-comparison}, together with
$1+2\sqrt{5\ln2}<5$ and $2\sqrt{\ln2}<2$, yields
\[
 \err_\mu(R)
 \le\frac{\err_\mu(R)}4
  +5\sqrt{\frac{L_{\mathrm{out}}}{\err_\mu(R)}t}
  +\frac{24(1+L_{\mathrm{in}})}{\sqrt R}t^{1/4}.
\]

The two terms in the minimum in~\eqref{eq:positive-contradiction}
now bound the two terms following $\err_\mu(R)/4$:
\[
 \begin{aligned}
 5\sqrt{\frac{L_{\mathrm{out}}}{\err_\mu(R)}t}
 &<5\cdot2^{-14}\err_\mu(R),\\
 \frac{24(1+L_{\mathrm{in}})}{\sqrt R}t^{1/4}
 &<24\cdot2^{-7}\err_\mu(R).
 \end{aligned}
\]
Consequently,
\[
 \err_\mu(R)
 <\left(\frac14+\frac5{2^{14}}+\frac{24}{2^7}\right)
   \err_\mu(R)
 <\err_\mu(R),
\]
a contradiction. This proves the positive-rate bound
in~\eqref{eq:explicit-positive}.

\proofstep{The case $R=0$.}
We use the zero-message protocol from
Lemma~\ref{lem:silent-coordinate}. Set $\varepsilon=\eta/3$
and suppose, for a contradiction, that
\[
 t<\frac{\eta^3}{128L_{\mathrm{out}}}.
\]
Since $\eta<1$ and $L_{\mathrm{out}}\ge1$,
$\delta(n,p)\le\sqrt t<1/2$, so the lemma applies.
Its protocol $\sigma$ has information cost zero, giving
$\eta=\err_\mu(0)\le\err_\mu(\sigma)$.
The error bound~\eqref{eq:silent-coordinate-error},
with $\varepsilon=\eta/3$, therefore yields
\begin{equation}
 \eta\le\frac\eta3+\delta(n,p)
 +2\sqrt{(\ln2)
 \left(2+\frac{L_{\mathrm{out}}}{-\log(1-\eta/3)}\right)t}.
 \label{eq:zero-specialization}
\end{equation}

The two terms after $\eta/3$ can be bounded together.
Since $-\log(1-\eta/3)\ge\eta/(3\ln2)$, we have
\[
 2+\frac{L_{\mathrm{out}}}{-\log(1-\eta/3)}
 \le2+\frac{3(\ln2)L_{\mathrm{out}}}{\eta}
 \le\frac{5L_{\mathrm{out}}}{\eta}.
\]
The last inequality uses $\eta<1$ and
$L_{\mathrm{out}}\ge1$. In addition,
$\delta(n,p)\le\sqrt t\le\sqrt{L_{\mathrm{out}}t/\eta}$.
Substituting these bounds into~\eqref{eq:zero-specialization}
and using $1+2\sqrt{5\ln2}<5$ gives
\[
 \eta\le\frac\eta3
      +5\sqrt{\frac{L_{\mathrm{out}}}{\eta}t}.
\]
The assumed bound on $t$ now implies
\[
 \eta<\left(\frac13+\frac5{\sqrt{128}}\right)\eta
 <\eta,
\]
a contradiction. Thus $t\ge\eta^3/(128L_{\mathrm{out}})$.

In either case, the established bound $t\ge\kappa(R)$
is equivalent to $p\le2^{-n\kappa(R)}$.
\end{proof}

\section{Consequences for communication complexity}
\label{sec:consequences}
Theorem~\ref{thm:converse} bounds the success probability at
every rate below $\IC_\mu(g,0)$ for a fixed input distribution
$\mu$. We combine it with amortized communication upper bounds
and then use the zero-error minimax identity~\cite{Bra17} to obtain the
randomized-communication statements. The upper bounds start from zero-error
protocols for one instance.

\begin{Prop}
\label{prop:amortized-upper-bounds}
For every $0<\varepsilon<1$ and every distribution $\mu$,
\begin{align}
 \limsup_{n\to\infty}
 \frac{D_{\mu^{\otimes n},\varepsilon}(g^n)}{n}
 &\le \IC_\mu(g,0),
 \label{eq:distributional-upper-bound}\\
 \limsup_{n\to\infty}
 \frac{R_\varepsilon(g^n)}{n}
 &\le \IC(g,0).
 \label{eq:prior-free-upper-bound}
\end{align}
\end{Prop}

\begin{proof}
The distributional bound~\eqref{eq:distributional-upper-bound}
follows from the interactive compression theorem of
Ref.~\cite{BR11}. While Ref.~\cite{BR11} focuses on per-instance error for computing $g^n$, its method provides a protocol $\widehat \pi_n$ that approximates $n$ independent executions of a fixed protocol $\pi$. Therefore if $\pi$ is zero-error, $\widehat \pi_n$ has small global error on $\mu^{\otimes n}$.

The zero-error randomized compression result of
Ref.~\cite{Sur26} gives
\eqref{eq:prior-free-upper-bound}.
\end{proof}

\subsection{Distributional communication}

\begin{Cor}
\label{thm:distributional-amortization}
For every distribution $\mu$ and every $0<\varepsilon<1$,
\begin{equation}
 \lim_{n\to\infty}
 \frac{D_{\mu^{\otimes n},\varepsilon}(g^n)}{n}
 =\IC_\mu(g,0)
 \label{eq:distributional-amortization}
\end{equation}
and an exponential strong converse holds for any rate below $\IC_\mu(g,0)$.
\end{Cor}

\begin{proof}
Proposition~\ref{prop:amortized-upper-bounds} gives the
upper bound in~\eqref{eq:distributional-amortization}.
For the lower bound, fix $0\le R<\IC_\mu(g,0)$.
Theorem~\ref{thm:converse} supplies a constant
$\kappa_\mu(R)>0$, independent of $n$ and of the protocol,
such that every protocol of worst-case communication cost at
most $nR$ succeeds under $\mu^{\otimes n}$ with probability at most
$2^{-n\kappa_\mu(R)}$. This is smaller than
$1-\varepsilon$ for all sufficiently large $n$. Hence
$D_{\mu^{\otimes n},\varepsilon}(g^n)>nR$ for those $n$.
Letting $R\uparrow\IC_\mu(g,0)$ proves the lower bound.
If $\IC_\mu(g,0)=0$, the lower bound is immediate.
\end{proof}

\subsection{Randomized communication}

We use the zero-error minimax identity~\cite{Bra17}
\begin{equation}
 \IC(g,0)=\max_\mu\IC_\mu(g,0).
 \label{eq:zero-error-minimax}
\end{equation}

\begin{Cor}
\label{cor:prior-free-strong-converse}
For every $0\le R<\IC(g,0)$ there are an input distribution
$\mu$ and a constant $\kappa>0$ such that every protocol
$\Pi_n$ for $g^n$ with $\Cost_{\mathrm{wc}}(\Pi_n)\le nR$
satisfies
\begin{equation}
 \min_{(x^n,y^n)}
 \Pr[\Pi_n\text{ succeeds on every coordinate}
      \mid x^n,y^n]
 \le 2^{-n\kappa}.
 \label{eq:prior-free-strong-converse}
\end{equation}
The distribution and exponent are independent of $n$ and
of $\Pi_n$.
\end{Cor}

\begin{proof}
The minimax identity~\eqref{eq:zero-error-minimax} gives a
distribution $\mu$ with $R<\IC_\mu(g,0)$.
Theorem~\ref{thm:converse} bounds the success probability of
$\Pi_n$ averaged over $(x^n,y^n)\sim\mu^{\otimes n}$ by
$2^{-n\kappa_\mu(R)}$. The minimum conditional success
probability over all input vectors is at most this average.
Taking $\kappa=\kappa_\mu(R)>0$ shows the statement. 
\end{proof}

The randomized-communication strong converse and
\eqref{eq:prior-free-upper-bound} also give the amortization
identity:
\begin{Cor}
\label{thm:prior-free-amortization}
For every $0<\varepsilon<1$,
\begin{equation}
 \lim_{n\to\infty}
 \frac{R_\varepsilon(g^n)}{n}
 =\IC(g,0).
 \label{eq:prior-free-amortization}
\end{equation}
\end{Cor}

\section*{Acknowledgement}
The author would like to thank Ashwin Nayak and Dave Touchette for their kindness and support and Atsuya Hasegawa and Masayuki Miyamoto for their monthly conversations. 
The author acknowledges the support of the Natural Sciences and Engineering Research Council of Canada (NSERC) [funding reference no. ALLRP-578455-2022]. This research was supported in part by the Institute for Quantum Computing.

\bibliography{Repetition_games_D, mathematics_D, Info_D, computational_D, query_D, quant_info_D, comm_comp_D, books_D}
\bibliographystyle{alpha}

\appendix
\section{Entropy bound and transcript factorization}
\label{app:facts}

We prove the entropy estimate used for the explicit exponent.
We also prove the transcript factorization used in
Appendix~\ref{app:zero-error}.

\subsection{An elementary bound for binary entropy}

\begin{proof}[Proof of Fact~\ref{fact:entropy-power}]
For $0<t<1$, the inequality $\ln u\le u-1$ with
$u=1/(1-t)$ gives
$(1-t)\ln(1/(1-t))\le t$.
Also $\sqrt t\ln(1/t)\le2/\exp(1)$: substituting
$u=\ln(1/t)\ge0$ reduces this to maximizing $u\exp(-u/2)$.
Consequently,
\[
 h(t)
 =\frac{t\ln(1/t)+(1-t)\ln(1/(1-t))}{\ln2}
 \le\frac{(2/\exp(1))\sqrt t+t}{\ln2}
 \le\frac{1+2/\exp(1)}{\ln2}\sqrt t
 <3\sqrt t.
\]
The endpoints follow from $h(0)=h(1)=0$.
\end{proof}

\subsection{Transcript factorization}

The next proposition uses the independence of the private
randomness in the protocol model of Section~\ref{sec:setting}.

\begin{Prop}
\label{prop:protocol-locality}
Fix a value $R_\textrm{pub}=r_0$ of positive probability and a transcript
$t$ of a finite protocol with independent private randomness.
There are nonnegative functions $u_{r_0,t}$ and $v_{r_0,t}$ such
that, whenever $(x,y,r_0)$ has positive probability,
\[
 \Pr[T=t\mid X=x,Y=y,R_0=r_0]
 =u_{r_0,t}(x)v_{r_0,t}(y).
\]
At every conditioning value of positive probability, the
outputs satisfy
\[
 P(A,B\mid X,Y,R_0,T)
 =P(A\mid X,R_0,T)P(B\mid Y,R_0,T).
\]
At any transcript prefix, the next message is conditionally
independent of the other party's input given the sender's input
and the public history.
\end{Prop}
\begin{proof}
For the fixed $r_0,t$, let $\chi_A(x,s_A)$ and $\chi_B(y,s_B)$
indicate whether each player's messages agree with $t$.
The transcript equals $t$ exactly when both indicators are one.
Independence of the private randomness gives
\[
 \Pr(S_A=s_A,S_B=s_B,T=t\mid x,y,r_0)
 =\Pr(S_A=s_A)\chi_A(x,s_A)
  \Pr(S_B=s_B)\chi_B(y,s_B).
\]
Summing over $s_A,s_B$ proves the transcript factorization.
At a conditioning value of positive probability, normalizing
the two factors separately shows that the private randomness
remains independent and each party's conditional distribution
depends only on its own input and $(r_0,t)$. The output
identity follows by applying the parties' output procedure.
\end{proof}

\providecommand{\muF}{\mu_{\mathrm{full}}}
\section{Continuity at zero error}
\label{app:zero-error}
\cref{prop:zero-error-continuity} shows the continuity of $\IC_\mu(g, \varepsilon)$ at $\varepsilon =0$, based on the argument given in Ref.~\cite[Section 6.2]{BGPW13} that shows the same continuity for relations with the same output set. The argument is slightly generalized to deal with relations with different output sets.

\begin{Prop}
\label{prop:zero-error-continuity}
For every total relation $g \subseteq \mathcal X \times \mathcal Y \times \mathcal A \times \mathcal B$ and
every input distribution $\mu$,
\[
 \IC_\mu(g,0)=\lim_{\varepsilon\downarrow0}
                         \IC_\mu(g,\varepsilon).
\]
\end{Prop}
\begin{proof}
The inequality $\IC_\mu(g,\varepsilon)\le\IC_\mu(g,0)$ follows
from the definition~\eqref{eq:def_IC_error}. For the
reverse inequality in the limit, fix $\pi$ with
$\err_\mu(\pi)\le\varepsilon$, where $\varepsilon>0$ is
sufficiently small. We repair $\pi$ to have zero error under $\mu$
at a cost tending to zero with $\varepsilon$.

We use a full-support distribution $\muF$ that converges to $\mu$ as $\varepsilon \downarrow 0$. 
Depending on $\muF$ a repaired protocol is constructed. 
The purpose of introducing $\muF$ is rather technical: it is introduced so that 
$\max_{(x,y) \in \supp(\muF)} u_t(x)v_t(y)= (\max_x u_t(x))(\max_y v_t(y))$ holds for some functions $u_t, v_t$. If not, we cannot in general obtain upper bounds on the functions defined below.

\medskip\noindent\emph{A full-support input distribution.}
 Write $m=|\mathcal X||\mathcal Y|$ and set
\[
\muF(x,y)=(1-\varepsilon^{1/8})\mu(x,y)
                         +\frac{\varepsilon^{1/8}}m.
\]
Thus $\muF(x,y)\ge\varepsilon^{1/8}/m$.
We also need to bound the error of $\pi$ under the full-support distribution $\muF$ on
$\supp(\mu)$. Since the positive values of $\mu(x,y)$ form a
finite set,
for all sufficiently small $\varepsilon$ we also have
$\varepsilon^{1/8}/m\le\mu(x,y)$ whenever $\mu(x,y)>0$.
On those inputs $\muF(x,y)\le2\mu(x,y)$, and hence
\begin{equation}
 \Pr_{\muF}\{\mu(X,Y)>0,\ \pi\text{ errs}\}\le2\varepsilon.
 \label{eq:continuity-restricted-error}
\end{equation}
\medskip\noindent\emph{Posterior distributions and fixed outputs.}

We now investigate a factorization property of $\muF$ conditioned on (shared randomness and) a transcript $t= (r_0, t_0)$. 
Fix a value $t=(r_0,t_0)$ of the public randomness $r_0$
and complete transcript $t_0$ of $\pi$ with positive probability. With a slight abuse of terminology, call $t$ a transcript as well.
The transcript factorization in Proposition~\ref{prop:protocol-locality} and
Bayes' rule give nonnegative functions $u_t,v_t$ such that
\begin{equation}
 \mu_t^{\mathrm{full}}(x,y):=
 \Pr_{\muF}(X=x,Y=y\mid R_\textrm{pub}=r_0,T=t_0)
 =u_t(x)v_t(y)\muF(x,y).
 \label{eq:continuity-rectangle}
\end{equation}
For later use, 
derive a uniform upper bound on these factors as follows. 
By rescaling without changing their product:
\begin{equation*}
v_t(y) \leftarrow \sqrt{\max_xu_t(x) / \max_yv_t(y)} v_t(y), \qquad
u_t(x) \leftarrow \sqrt{\max_xu_t(x) / \max_yv_t(y)}^{-1} u_t(x), 
\end{equation*}
it holds that 
\begin{equation*}
	\max_xu_t(x)=\max_yv_t(y) = \sqrt{\max_xu_t(x) \cdot \max_yv_t(y)}.
\end{equation*}
At an input pair $(x,y)$ where both maxima occur, the expression~\eqref{eq:continuity-rectangle} tells 
\[
 1\ge\mu_t^{\mathrm{full}}(x,y)
 \ge\frac{\varepsilon^{1/8}}m
        \bigl(\max_xu_t(x)\bigr)^2.
\]
Consequently,
\begin{equation}
 \max_xu_t(x)=\max_yv_t(y)
       \le\sqrt m\,\varepsilon^{-1/16},
 \label{eq:continuity-factor-max}
\end{equation}
and this is the purpose of introducing $\muF$.

For our analysis, the parties should output deterministically once the protocol is performed, while in general they may use private randomness for outputs.
To this end, we will fix the outputs $(a_t(x), b_t(y)) \in \mathcal A \times \mathcal B$ of the parties that depend only on the inputs and $t=(r_0, t_0)$, while preserving the conditional error on $\supp(\mu)$. 

Conditional on $(X,Y,R_\textrm{pub},T)=(x,y,r_0,t_0)$, the players'
output distributions depend only on $(x,r_0,t_0)$ and $(y,r_0,t_0)$,
respectively, and are independent, by
Proposition~\ref{prop:protocol-locality}. 
Sample an output $a_t(x)$ from Alice's conditional output
distribution for every $x$, and independently sample $b_t(y)$ from
Bob's for every $y$. The expected error mass of these randomized outputs on
$\supp(\mu)$ is the error of $\pi$ conditional on $t$. Thus we can fix
deterministic functions $a_t, b_t$ for which
\begin{equation}
 e_t:=
 \sum_{\mu(x,y)>0}\mu_t^{\mathrm{full}}(x,y)
       \mathbf1\{(a_t(x),b_t(y))\notin g(x,y)\}
 \le
 \Pr_{\muF}\{\mu(X,Y)>0,\ \pi\text{ errs}\mid (R_\textrm{pub},T)=t\}.
 \label{eq:continuity-fixed-outputs}
\end{equation}
The error bound~\eqref{eq:continuity-restricted-error} gives
$\mathbb E_{\muF}e_{(R_\textrm{pub},T)}\le2\varepsilon$. Hence Markov's
inequality gives
\[
	\Pr_{\muF}\{e_{(R_\textrm{Pub},T)}>\sqrt{2\varepsilon}\}
 \le\sqrt{2\varepsilon}.
\]

\medskip\noindent\emph{Correctness on $L_A(t) \times L_B(t)$.}
We show that when $e_t\le\sqrt{2\varepsilon}$ and both $u_t$ and $v_t$ exceed
$\varepsilon^{1/8}$,
 the deterministic functions $a_t, b_t$ must output correctly at every input in $\supp(\mu)$.
Indeed, if $\mu(x,y)>0$
and both $u_t(x)>\varepsilon^{1/8}$ and
$v_t(y)>\varepsilon^{1/8}$ for an input pair $(x,y)$, then
\[
 \mu_t^{\mathrm{full}}(x,y)
 >\varepsilon^{1/4}\muF(x,y)
 \ge\frac{\varepsilon^{3/8}}m
 >\sqrt{2\varepsilon},
\]
where the last inequality holds once
$\varepsilon^{1/8}<1/(\sqrt2m)$. 
If the fixed outputs were incorrect at $(x,y)$, its contribution to
$e_t$ in~\eqref{eq:continuity-fixed-outputs} would exceed
$\sqrt{2\varepsilon}$, contradicting the assumption $e_t \le \sqrt{2\varepsilon}$. Thus
$(a_t(x),b_t(y))\in g(x,y)$ whenever both inequalities hold
and $\mu(x,y)>0$ and $e_t\le\sqrt{2\varepsilon}$.

We next evaluate how large the sets
\begin{equation*}
	L_A(t) := \{x : u_t(x) > \varepsilon^{1/8}\}, \qquad
	L_B(t) := \{y : v_t(y) > \varepsilon^{1/8}\}
\end{equation*}
are. 
Equations~\eqref{eq:continuity-rectangle} and
\eqref{eq:continuity-factor-max} give
\[
 \begin{aligned}
		 \mu_t^{\mathrm{full}}(L_A(t)^c)
 = \mu_t^{\mathrm{full}}\{u_t(X)\le\varepsilon^{1/8}\}
 &\le\sqrt m\,\varepsilon^{1/16}
      \sum_{\substack{x:u_t(x)\le\varepsilon^{1/8}\\y}}
      \muF(x,y)
 \le\sqrt m\,\varepsilon^{1/16},\\
 \mu_t^{\mathrm{full}}(L_B(t)^c)
 = \mu_t^{\mathrm{full}}\{v_t(Y)\le\varepsilon^{1/8}\}
 &\le\sqrt m\,\varepsilon^{1/16}
 \end{aligned}
\]
when $\Pr_{\muF}\{(R_\textrm{Pub},T)=t\}>0$.
Thus, for every transcript $t$ with positive probability,
\begin{equation}
		1- \mu_t^{\mathrm{full}}(L_A(t) \times L_B(t))
 = \mu_t^{\mathrm{full}}\{u_t(X)\le\varepsilon^{1/8}
             \text{ or }v_t(Y)\le\varepsilon^{1/8}\}
 \le2\sqrt m\,\varepsilon^{1/16}.
 \label{eq:continuity-rectangle-mass}
\end{equation}

\medskip\noindent\emph{Zero-error repair and its information cost.}
We now construct a repaired protocol $\widehat\pi$ with zero error
under $\mu$.
\begin{enumerate}
 \item Run $\pi$ and let $t=(r_0,t_0)$.
 \item If $e_t>\sqrt{2\varepsilon}$, exchange the inputs.
 \item Otherwise, each party sends one bit to indicate whether
 $x\in L_A(t)$ or $y\in L_B(t)$, respectively. If both bits are
 one, output $(a_t(x),b_t(y))$. If either bit is zero, exchange
 the inputs and choose a valid output pair for $g(x,y)$.
\end{enumerate}
To check $\widehat \pi$ is zero error, take any $(x,y)$ with $\mu(x,y)>0$. If the
inputs are exchanged, both parties can choose a correct output pair.
 Otherwise $e_t\le\sqrt{2\varepsilon}$ and
$(x,y)\in L_A(t)\times L_B(t)$. The posterior inequality above
gives $\mu_t^{\mathrm{full}}(x,y)>\sqrt{2\varepsilon}$, so an
incorrect output $(a_t(x),b_t(y))$ would contribute more than
$\sqrt{2\varepsilon}$ to $e_t$. Therefore $\widehat\pi$ has zero
error under $\mu$.

Let $F$ be the event that the inputs are exchanged, and let $U$
consist only of the additional communicated bits. Set
\[
 \delta:=\sqrt{2\varepsilon}
                   +2\sqrt m\,\varepsilon^{1/16},
\]
which tends to zero with $\varepsilon$. The Markov bound and the
size~\eqref{eq:continuity-rectangle-mass} of $L_A(t) \times L_B(t)$ give
\begin{equation*}
\Pr_{\muF}(F)\le\sqrt{2\varepsilon}+2\sqrt m\,\varepsilon^{1/16}.
\end{equation*}
 Since
$\muF=(1-\varepsilon^{1/8})\mu+
\varepsilon^{1/8}\operatorname{Unif}(\mathcal X\times\mathcal Y)$,
the same protocol, including its random choices, satisfies
\[
 \Pr_\mu(F)\le
 \frac{\Pr_{\muF}(F)}{1-\varepsilon^{1/8}}
 \le \frac{\sqrt{2\varepsilon}+2\sqrt m\,\varepsilon^{1/16}}{1-\varepsilon^{1/8}}
 =:\beta,
\]
where $\beta$ vanishes as $\varepsilon \downarrow 0$.
When $F$ does not occur, $U$ is the fixed pair of membership bits
$11$. When $F$ occurs, $U$ contains at most two membership bits
and encodings of both inputs, using at most $2+L_{\mathrm{in}}$
bits. For sufficiently small $\varepsilon$ we have $\beta\le1/2$.
Conditioning on $F$ and using the entropy chain rule therefore
gives
\[
\begin{aligned}
H_\mu(U\mid R_\textrm{Pub},T)
&\le H_\mu(F,U\mid R_\textrm{Pub},T)\\
&= H_\mu(F\mid R_\textrm{Pub},T)
   +H_\mu(U\mid F,R_\textrm{Pub},T)\\
&= H_\mu(F\mid R_\textrm{Pub},T)
   +\Pr_\mu(F^c)H_\mu(U\mid R_\textrm{Pub},T,F^c)\\
&\quad+\Pr_\mu(F)H_\mu(U\mid R_\textrm{Pub},T,F)\\
&\le h(\Pr_\mu(F))
   +\Pr_\mu(F)\log(4|\mathcal X||\mathcal Y|)\\
&= h(\Pr_\mu(F))
   +\Pr_\mu(F)(2+L_{\mathrm{in}})\\
&\le h(\beta)+\beta(2+L_{\mathrm{in}}).
\end{aligned}
\]
The chain rule for the information cost of $\widehat\pi$ gives
\begin{equation}
 \begin{aligned}
 \IC_\mu(\widehat\pi)-\IC_\mu(\pi)
 &=I_\mu(X;U\mid Y,R_\textrm{Pub},T)
 +I_\mu(Y;U\mid X,R_\textrm{Pub},T)\\
 &\le 2H_\mu(U\mid R_\textrm{Pub},T)\\
 &\le 2h(\beta)+2\beta(2+L_{\mathrm{in}}).
 \end{aligned}
 \label{eq:continuity-repair-cost}
\end{equation}
Since $\widehat\pi$ has zero error under $\mu$,
\[
 \IC_\mu(g,0)\le\IC_\mu(\widehat\pi)
 \le\IC_\mu(\pi)+2h(\beta)+2\beta(2+L_{\mathrm{in}}).
\]
The added term is independent of $\pi$ and tends to zero with
$\varepsilon$. Taking the infimum over protocols with error at
most $\varepsilon$, then letting $\varepsilon\downarrow0$,
proves the reverse inequality in the statement.
\end{proof}

\section{Conditional sampling}
\label{app:conditional-sampling}
\cref{lem:shared-sampling} gives a standard sampling procedure based on rejection sampling argument.

\begin{Lem}
\label{lem:shared-sampling}
Let $\mathcal D$ be a finite family of distributions on a
finite set $\mathcal V$. There are a finite-valued random
variable $R$ and functions $S_D(R)\in\mathcal V$, for
$D\in\mathcal D$, such that $S_D(R)\sim D$ and
\begin{equation}
 \Pr[S_D(R)\ne S_E(R)]
 \le \frac{2\TV(D,E)}{1+\TV(D,E)}
 \le 2\TV(D,E)
 \quad(D,E\in\mathcal D).
 \label{eq:shared-sampling-bound}
\end{equation}
\end{Lem}

\begin{proof}
Write $m=|\mathcal V|$. For each trial $j= 1, 2, \ldots$, independently
choose $Z_j$ uniformly from $\mathcal V$ and $U_j$
uniformly from $[0,1]$. For a distribution $D$, accept
$Z_j$ when $U_j\le D(Z_j)$, and return the value in the
first accepted trial.
One trial accepts $z$ under $D$ with probability $\Pr(z) \Pr(\{U_j \le D(z)\}\mid z) = D(z)/m$,
and accepts some value with probability $1/m$. Thus the
probability of returning $z$ in the first $N$ trials is
\begin{equation}\label{eq:app:lem:con}
 \sum_{j=1}^N (1-1/m)^{j-1}\frac{D(z)}m
 =D(z)\bigl[1-(1-1/m)^N\bigr].
\end{equation}
For $D,E\in\mathcal D$, put $t=\TV(D,E)$. In one trial,
the probability that at least one of $D$ and $E$ accepts is, from $\max\{a, b\} = (a + b + |a - b|)/2$,
\[
	\Pr\left[\exists z : Z_j =z \Mand U_j \le \max\{D(z),E(z)\}\right]
=  \frac1m\sum_z\max\{D(z),E(z)\}
 =\frac{1+t}{m},
\]
and the probability that exactly one accepts is
\begin{align*}
	&\Pr\bigl[\exists z : Z_j =z \Mand  \min \{D(z), E(z) \} < U_j \le \max\{D(z),E(z)\}\bigr]\\
	&= \frac1m\sum_z|D(z)-E(z)|=\frac{2t}{m}.
\end{align*}
Therefore the probability
that they both finish within $N$ trials and return different
values is at most
\begin{align}
	&\Pr\bigl[\text{One of $E$ or $D$ finishes earlier within $N$ trials}\bigr] \notag \\
	&\le
 \sum_{j=1}^N
 \left(1-\frac{1+t}{m}\right)^{j-1}\frac{2t}{m}
 = \biggl[1 - \left(1 - \frac{1+t}{m} \right)^N \biggr]\cdot \frac{m}{1+t} \frac{2t}{m} \label{align:lem:app:con}
	\le \frac{2t}{1+t}.
\end{align}
Together with \eqref{eq:app:lem:con}, the infinite sequence
$(Z_j,U_j)_{j\ge1}$ gives shared randomness satisfying
the requirements. It remains
to replace this sequence by a finite-valued shared random
variable.

For each $D\in\mathcal D$, let
\[
 T_D=\min\{j\ge1:U_j\le D(Z_j)\}.
\]
For $s = (z_D)_{D \in \mathcal D}\in\mathcal V^{\mathcal D}$, define
\[
 p_N(s):=
 \Pr[T_D\le N,\ Z_{T_D}=z_D
       \text{ for every }D\in\mathcal D],
 \qquad
 p(s)=\lim_{N\to\infty}p_N(s).
\]
The limit exists because $p_N(s)$ is nondecreasing in
$N$. Moreover,
\[
 1-\sum_s p_N(s)
 =\Pr[\text{some }T_D>N]
 \le |\mathcal D|(1-1/m)^N.
\]
As $\mathcal V^{\mathcal D}$ is finite, $\sum_s p(s)=1$.
Thus we may take a random variable $R$ on the finite
set $\mathcal V^{\mathcal D}$ with
\[
 \Pr[R=s]=p(s)=\lim_{N\to\infty}
 \Pr[T_D\le N,\ Z_{T_D}=z_D
       \text{ for every }D\in\mathcal D],
\]
and define $S_D(s)=z_D$. In fact, the same bound gives
\[
 0\le p(s)-p_N(s)\le|\mathcal D|(1-1/m)^N
 \qquad(s\in\mathcal V^{\mathcal D}).
\]
For $D\in\mathcal D$ and $z\in\mathcal V$,
\[
 \Pr[S_D(R)=z]
 =\lim_{N\to\infty}\Pr[T_D\le N,\ Z_{T_D}=z]
 =D(z)
\]
by \eqref{eq:app:lem:con}. Similarly, for $D,E\in\mathcal D$,
the finite-trial bound~\eqref{align:lem:app:con} above gives
\[
 \Pr[S_D(R)\ne S_E(R)]
 =\lim_{N\to\infty}
   \Pr[T_D,T_E\le N,\ Z_{T_D}\ne Z_{T_E}]
 \le \frac{2\TV(D,E)}{1+\TV(D,E)}.
\]
This proves \eqref{eq:shared-sampling-bound}.
\end{proof}

\end{document}